\documentclass[epj]{svjour}
\usepackage{graphics}
\usepackage{amssymb}
\usepackage{amsmath}
\usepackage{cite}
\usepackage{multirow}
\usepackage{multicol}
\usepackage{gensymb}
\usepackage{longtable}
\usepackage{booktabs}
\usepackage{array}
\usepackage{anyfontsize} 
\usepackage{ulem}
\usepackage{wrapfig}
\usepackage{xcolor}
\usepackage{cancel}
\usepackage{indentfirst}
\definecolor{tit}{rgb}{0.1,0.2,0.4}

\newcommand{\beq}{\begin{eqnarray}}
\newcommand{\eeq}{\end{eqnarray}}

\renewcommand{\sl}[1]{#1 \hspace{-0.2cm}/}

\newcommand{\bea}{\begin{eqnarray}}
\newcommand{\eea}{\end{eqnarray}}
\newcommand{\be}{\begin{equation}}
\newcommand{\ee}{\end{equation}}
\newcommand{\non}{\nonumber\\ }
\newcommand{\vsl}{ v \hspace{-2.2truemm}/ }

\newcommand{\xsl}{ x \hspace{-2.2truemm}/ }
\newcommand{\psl}{ p \hspace{-2.2truemm}/ }

\definecolor{DRed}{rgb}{0.8,0,0.1}
\definecolor{DBlue}{rgb}{0,0,0.8}

\begin{document}
\title{$\bar{B}_s \to f_0(980)$ form factors and the width effect from light-cone sum rules }
\author{Shan Cheng%\thanks{ }%\inst{2}
\and Jian-Ming Shen %\inst{2}
% etc
}                     % Do not remove
\mail{Shan Cheng}
\institute{School of Physics and Electronics, Hunan University, 
410082 Changsha, People's Republic of China, \non 
\email{scheng@hnu.edu.cn, shenjm@hnu.edu.cn}.
%\and other institute
}
\date{Received: date / Revised version: date}
\abstract{
In this paper we calculate the $\bar{B}_s \to f_0(980)$ form factors from light-cone sum rules with $B$ meson DAs. 
With adopting the quark-antiquark configuration of light scalar mesons, 
the high twist two-particle and the three-particle contributions are found to be $\sim 25\%$ individual, 
and totally they give about $50\%$ correction to certain form factors in the considered energy regions. 
We further explore the light-cone sum rules approach to study the $S-$wave $\bar{B}_s \to KK$ form factors, 
the $f_0+f'_0+f''_0$ resonance model is proposed and the result shows that 
the background effect from $f'_0+f''_0$ accounts $\sim 5\%$. 
As a by-product, we extract the strong coupling $|g_{f_0 KK }| = 1.08^{+0.05}_{-0.14}$ GeV 
with taking the $\bar{B}_s \to f_0(980)$ form factors calculated previous under the narrow width approximation.
 % end of PACS codes
              } 
%end of abstract

\maketitle
\section{Introduction} \label{sec:introduction}

Form factor is a fundamental physical quantum in effective field theory (EFT), 
with including both the long distance (LD) and short distance (SD) physics. 
In order to extract the Cabibbo-Kobayashi-Maskawa ($\text{CKM}$) matrix elements, reliably, in the semileptonic $B$ decay processes, 
the precise calculations of the relevant form factors are inevitable.
The heavy-to-light form factors are calculated by different approaches, 
in which the lattice QCD (LQCD) gives the reliable simulations in the low recoiled regions \cite{HorganPVA,LatticeTIA}, 
while in the large and full recoiled regions, the QCD-based analytical approaches like the light-cone sum rules (LCSRs) \cite{BraunIJ,BraunKW,ColangeloDP,BallYE,BallRG,StraubICA,KhodjamirianST,DuplancicIX,WangVGV,ChengSMJ,Descotes-GenonBUD,RusovCHR} 
and the perturbative QCD (PQCD) \cite{LiNU,LiIU,KurimotoZJ,LuNY,ChenGV,LiNK,ChengFWA} are applied. 

There are many successful calculations for the heavy-to-light transition form factors 
with the final state being a pseudoscalar ($P$) or a vector ($V$) meson. 
While to our knowledge, the transition form factors with light scalar ($S$) meson final state are not understood well so far
due to the unclear underlying structure and the large width effect. 
It has been suggested that the scalar mesons with masses below or near $1$ GeV (the isoscalar $\sigma/f_0(500)$ and $f_0(980)$, 
the isodoublet $\kappa$, and the isovector $a_0$) form a $SU(3)$ flavor nonet, 
and the scalar mesons with masses around $1.5$ GeV ($f_0(1370)$, $a_0(1450)$, $K_0^\ast(1430)$ and $f_0(1500)$) form the other one. 
The combined analysis based on orbital angular momentum \cite{JaffeIG,ChengNB} and data \cite{CloseZU,AchasovHM,AchasovFH} 
implies that the heavier nonet states favour the quark-antiquark configuration replenished with some possible gluon content. 
From the spectral analysis, 
there is not a general agreement on the underlying assignments of the scalar mesons in the lighter nonet, like $f_0(980)$. 
Pictures like tetra-quark \cite{JaffeIG,Agaev:2017cfz,Agaev:2018sco}, 
gluonball \cite{WeinsteinGC}, hybrid state \cite{WeinsteinGD} and molecule state \cite{WeinsteinGU} are all discussed, 
in which the tetra-quark assignment is more favorite nowadays. 
The case is different in the $B \to f_0(980)$ decays when $f_0(980)$ is energetic and the process happens with large recoiling, 
where the conventional quark-antiquark assignment is the favorite one 
since the possibility to form a tetra-quark state is power suppressed with comparing to the state of quark pair \cite{ChengNB}. 
So in this work, with the main purpose to calculate the heavy-quark transition form factors, 
we would take the usual quark-antiquark nature of $f_0(980)$, 
and postpone the tetra-quark study somewhere else.

Some attempts are carried out to calculate the $B \to S$ transition form factors 
from LCSRs with scalar mesons distribution amplitudes (DAs) \cite{GhahramanyZZ,SunNV,ColangeloBG,WangVRA}. 
We comment that these work considered only the scalar mesons in the heavier nonet, 
and their accuracy is debatable since some important informations of the input DAs, 
such as the standard conformal partial expansion and the width effect, are still missing. 
In this paper, with taking the $\bar{s}s$ configuration of $f_0(980)$, 
we suggest to study the $\bar{B}_s \to f_0(980)$ form factor from the alternative LCSRs with $B$ meson DAs. 
Although the width of $f_0(980)$ is suppressed by the phase space\footnote{In fact 
the width of $f_0$ is smaller than it of $\rho$ meson. 
It should be stressed that the width effect in $B\to\rho$ transition is usually neglected 
because in the experimental analysis the $\rho$ meson is identified by the $\text{P}$-wave $\pi\pi$ signal 
when the dipion invariant mass locates in the $\rho-$pole region \cite{BehrensVV,AdamPV}. 
This makes the narrow-width treatment for $B\to\rho$ form factor from LCSRs being consistent with the experimental measurement.}, 
we would like to access the width effect by applying the approach proposed to 
calculate $B \to \pi\pi, \, K\pi$ form factors \cite{ChengSMJ,Descotes-GenonBUD}, 
with substituting the isovector $\pi\pi$ state by the scalar isoscalar $KK$ state.

The rest of this paper is organised as follows. 
In section \ref{QCDSR} we revisit and update the mass and the decay constant of $f_0(980)$ in the two-point QCD sum rules. 
Section \ref{LCSRs} and section \ref{width-effect} are the main parts of this paper, 
where we present the LCSRs calculation for $\bar{B}_s \to f_0(980)$ form factors and 
generalizes it to study the $S-$wave $\bar{B}_s \to KK$ form factors, respectively. 
We summary in section \ref{conclusion}. 
The coefficients in three-particle corrections from $B$ meson LCDAs are complemented in Appendix \ref{app:coeff-3p}.

\section{$\sigma$ and $f_0(980)$ in the QCD sum rules revisited}\label{QCDSR}

QCD sum rules \cite{ShifmanBX} is a powerful tool to study hadron spectrum. 
For the meson with $I^G(J^{PC}) = 0^+(0^{++})$, 
the scalar isoscalar currents include both $J^n=\bar{n}n=\frac{1}{\sqrt{2}}(\bar{u}u+\bar{d}d)$ and $J^s=\bar{s}s$. 
We start with the two-point correlation function  
{\small
\beq
\Pi_{\text{2pSRs}}(q) = i \int d^4x e^{i q x} \langle 0 |T \{J^s(x),J^s(0)\} | 0 \rangle \,. 
\label{eq:corr-2psr}
\eeq} 
Although the data of $D_s^+ \to f_0 \pi^+$ indicates $f_0(980)$ may be dominated by the $\bar{s}s$ component, 
much more measurements \cite{GokalpNY,AnisovichZP,ChengAI,KaminskiQG,MennessierXG,ZhangKM,LiSW,ZhangQVQ,LiuYMI,FeldmannSH}, 
especially the comparable branching ratios between $B \to f_0(980) \to \pi\pi$ and $B \to f_0(980) \to KK$ \cite{TanabashiOCA}, 
support a mixing between $f_0$ and $\sigma$:  
{\small
\beq
&&| f_0(980) \, \rangle = | s\bar{s} \, \rangle \cos \theta + | n\bar{n} \, \rangle \sin \theta \,,\non
&&| \sigma(500) \, \rangle = - | s\bar{s} \, \rangle \sin \theta + | n\bar{n} \, \rangle \cos \theta \,.
\label{eq:mixing}
\eeq}

The mixing implies that $f_0$ and $\sigma$ should be treated separately, 
and in the basis of flavour, two decay constants  are needed to describe each of them,
{\small
\beq
&&\langle f_0 \vert \bar{u}u \vert 0 \rangle = \frac{1}{\sqrt{2}} m_{f_0} \bar{f}_{f_0}^n \, , \,\,\,\,\,\,
\langle f_0 \vert \bar{s}s \vert 0 \rangle =  m_{f_0} \bar{f}_{f_0}^s \, , \non
&&\langle \sigma \vert  \bar{u}u \vert 0 \rangle = \frac{1}{\sqrt{2}} m_{\sigma} \bar{f}_{\sigma}^n \, , \,\,\,\,\,\,\,\,\,\,
\langle \sigma \vert \bar{s}s \vert 0 \rangle =  m_{\sigma} \bar{f}_{\sigma}^s \, .
\label{eq:m-f-scalar}
\eeq}
The neutral scalar meson can not be produced via the vector current 
because $f_S$ is vanished in the $SU(3)$/isospin limit with the charge conjugation invariance and the conservation of vector current. 

The basic idea of QCD sum rules is to calculate independently for the correlation function in twofold ways: 
the QCD calculation at quark-gluon level in the Euclidean momenta space, 
and the summing of intermediate states from the view of hadron. 
The QCD calculation in the negative half plane of $q^2$ is guaranteed by the operator-product-expansion (OPE) technology, 
and the correlation function is then written in terms of various vacuum condensates. 
On the other hand, the average distance between two coordinate points ($0$ and $x$ in Eq.\ref{eq:corr-2psr}) grows 
when $q^2$ shifting from large negative to positive values, 
and then the LD quark-gluon interaction forms the hadrons \cite{ColangeloDP}. 
In this way, the correlation function can be expressed as the sum of contributions from all possible 
intermediate states in the positive half-plane, with possible subtractions. 
The accuracy of LCSRs approach is mainly depended on how to match the QCD calculation to the hadron spectral analysis, 
in more word is how to take the quark-hadron duality in the dispersion relation to eliminate the contributions from excited and continuum states. 

By inserting a complete set of intermediate states $\vert n \rangle$, the unitarity relation ($q^2 > 0$) of the correlation function reads as, 
{\small
\beq
&~& 2 \, \textrm{Im} \Pi_{\mathrm{2pSRs}} (q) \non
&=& \sum_n \, \langle 0 \vert J^s(x) \vert n \rangle \langle n \vert J^s(0) \vert 0 \rangle \, d \tau_n \, (2 \pi)^4 \, \delta(q-p_n) \,,
\label{eq:hadron-unitary}
\eeq}
where $d \tau_n$ denotes the phase space of each state $\vert n \rangle$, like $\sigma\,, f_0(980)$ and their excited states. 
We are now interesting in $f_0(980)$ which enters in the two-point sum rules as the first excited state with the ground state $\sigma$, 
so we include both them in the hadron inserting and take the threshold $s_0$ to truncate the higher excited states. 
After applying the dispersion relation and employing the quark-hadron duality, 
the matching between quark amplitude and the hadron spectral analysis is taken as 
{\small
\beq
\sum_{S=\sigma,f_0} \frac{m_S^2 (\bar{f}_{S}^s)^2}{(m_S^2-q^2)} 
= \frac{1}{\pi} \int_{0}^{s_0} ds \frac{\text{Im} \, \Pi^{\text{OPE}}_{\text{2pSRs}}(s)}{s-q^2} \,,
\label{eq:sr-0}
\eeq}
where $\Pi^{\text{OPE}}_{\text{2pSRs}}(s)$ is the OPE result for the correlation function. 
%Actually, the large width of $\sigma$ violates the spectral formula in the hadron interpolating, which problem is unable to be solved in the traditional QCD sum rules approach and exceed the scope of this work, so we adopt the coarse-grained approximation with narrow widths for $\sigma$.  
In order to improve the convergence of OPE calculation and suppress the contributions from high excited states and continuum spectrums, 
we apply the Borel transformation to both sides of Eq.\ref{eq:sr-0}. 
The result is quoted \cite{GovaertsUA,BecchiVZ} as follow, 
with $\alpha_s$ to one-loop order and the vacuum condensate terms up to dimension six, 
{\small
\beq
&~& \sum_{S=\sigma,f_0} m_S^2 (\bar{f}_S^s(\mu_0))^2 e^{-m_S^2/M^2} 
\Big(\frac{\alpha_s(\mu_0)}{\alpha_s(M)}\Big)^{2/\beta_1} \non
&=& 
\frac{3}{8\pi^2} M^4 \Big[ 1 +h(1) \Big]  f(1) + \frac{1}{8} \langle \frac{\alpha_s G^2}{\pi} \rangle + 3 m_s \langle \bar{s}s \rangle \non
&-& \frac{1}{M^2}\Big[ m_s \langle \bar{s} g_s \sigma \cdot G s \rangle 
-\frac{2}{3} \pi \alpha_s \langle \bar{s} \gamma_\mu \lambda^a s \bar{u} \gamma^\mu \lambda^a s \rangle \non
&-& \pi \alpha_s \langle \bar{s} \sigma_{\mu\nu} \lambda^a s \bar{s} \sigma^{\mu\nu} \lambda^a s \rangle \Big] \,. 
\label{eq:scalar-sr}
\eeq}
The functions defined in the perturbative terms are 
{\small
\beq
&&f(n) = 1 - e^{-\frac{s_0}{M^2}} \Big[1+\frac{s_0}{M^2}+\frac{1}{2} \Big(\frac{s_0}{M^2}\Big)^2 + \frac{1}{n !} \Big(\frac{s_0}{M^2}\Big)^n\big] \,,\non
&&I(n) = \int_{e^{-\frac{s_0}{M^2}}}^{1}\, dt \, \ln^n t \, \ln(-\ln t) \,, \non
&&h(n) = \frac{\alpha_s(M)}{\pi}\Big(\frac{17}{3} + 2 \frac{I(1)}{f(1)} - 2 \ln \frac{M^2}{\mu^2} \Big) \,.
\label{eq:scalar-sr-pert-1}
\eeq}

We use the two-loop expression for strong coupling \cite{TanabashiOCA}, 
{\small
\beq
\alpha_s(\mu) = \frac{\pi}{2 \beta_1 \log(\mu/\Lambda)} 
\Big(1-\frac{\beta_2}{\beta_1^2}\frac{\log(2\log(\mu/\Lambda))}{2 \log(\mu/\Lambda)} \Big) \,, 
\eeq}
with the evolution kernels $\beta_1 = (33 - 2 n_f)/12$ and $\beta_2 = (153 - 19n_f)/24$. 
The hadronic scale $\Lambda$ is set to reproduce $\alpha_s(m_Z) = 0.118$ and $\alpha_s(1 \, \mathrm{GeV}) = 0.474$, 
and written in terms of step function  
{\small
\beq
\Lambda^{(n_f)} = \mathrm{Which} \, [n_f=3, \, 0.332, \, n_f=4, \, 0.292, \, n_f=5, \, 0.210] \,.
\eeq}
The input values for the nonperturbative vacuum condensates at default scale $1 \, \mathrm{GeV}$ are taken as \cite{LeutwylerQG,IoffeEE}: 
{\small
\beq
&&\langle \bar{u}u \rangle = (-0.25 \, \mathrm{GeV})^3 \,,  \,\,\,\,\,\, 
\langle \bar{s}s \rangle = 0.8 \langle \bar{u}u \rangle \,, \non
&&\langle g_s \bar{q} \sigma \cdot G q \rangle = -0.8 \langle \bar{q}q \rangle \,, \non
&&\langle \alpha_s/\pi G^a_{\mu\nu}G^{a \mu\nu} \rangle = 0.012 \, \mathrm{GeV}^4 \,, \non
&&\langle \bar{q} \gamma_\mu \lambda^a q \bar{q} \gamma^\mu \lambda^a q \rangle = - \frac{16}{9} \langle \bar{q} q \rangle^2\,, \non
&&\langle \bar{q} \sigma_{\mu\nu} \lambda^a q \bar{q} \sigma^{\mu\nu}  \lambda^a q \rangle = - \frac{3}{4} \langle \bar{q} q \rangle^2\,.
\label{eq:vc}
\eeq}
The light quark masses are estimated as the "current-quark" masses 
in the $\overline{\mathrm{MS}}$ ($\mu = 1 \, \mathrm{GeV}$) scheme \cite{TanabashiOCA},
{\small
\beq
%m_u \equiv \frac{m_u + m_d}{2}= 0.0045 \, \mathrm{GeV}\,,\,\,\, 
m_s = 0.125\, \mathrm{GeV} \,.
\label{eq:quark-mass}
\eeq}
We also consider the running of parameters with one-loop accuracy \cite{ChernyakAS,LepageZB,VermaserenFQ},
{\small
\beq
&&m_{q}(\mu) = m_q(\mu_0) \Big(\frac{\alpha_s(\mu_0)}{\alpha_s(\mu)}\Big)^{-1/\beta_1} \,, \non
&&\langle \bar{q}q \rangle_\mu = \langle \bar{q}q \rangle_{\mu_0} \Big(\frac{\alpha_s(\mu_0)}{\alpha_s(\mu)}\Big)^{1/\beta_1} \,, \non
&&\langle \alpha_s G^2 \rangle_\mu = \langle \alpha_s G^2 \rangle_{\mu_0} \,, \non
&&\langle g_s \bar{q} \sigma \cdot G q \rangle_\mu = \langle g_s \bar{q} \sigma \cdot G q\rangle_{\mu_0} \Big(\frac{\alpha_s(\mu_0)}{\alpha_s(\mu)} \Big)^{-1/(6\beta_1)} \,. 
\label{eq:scalar-sr-nonpert}
\eeq}

Differentiating both sides of Eq.\ref{eq:scalar-sr} by the Borel mass we obtain an auxiliary sum rules, 
%\columnseprule=1pt
{\small
\beq
&~& \sum_{S=\sigma,f_0} \frac{m_S^4 (\bar{f}_S^s(\mu_0))^2 e^{-m_S^2/M^2} }{m_S^2 (\bar{f}_S^s(\mu_0))^2 e^{-m_S^2/M^2} } \non 
&=& \frac{ 2 M^2 \Big[ 1 +  h(2) \Big]  f(2)
+ \frac{8\pi^2}{3 } \frac{68\pi \alpha_s}{27 M^4}  \langle \bar{s}s \rangle^2}
{\Big[ 1 + h(1) \Big]  f(1)
+ \frac{8\pi^2}{3} \left[ \frac{1}{8 M^4} \langle \frac{\alpha_s G^2}{\pi} \rangle- \frac{68\pi \alpha_s}{27 M^6}  \langle \bar{s}s \rangle^2 \right]} \,, 
\label{eq:scalar-sr-1}
\eeq }
which is further used, together with the renormalization-improved sum rules in Eq.\ref{eq:scalar-sr}, 
to fit the masses and decay constants of $\sigma$ and $f_0$.
The terms proportional to quark mass on the right hand side are neglected in Eq.\ref{eq:scalar-sr-1}. 

The Borel mass is fixed by the rule of thumb that 
the contribution from high dimension condensate terms is no larger than twenty percents in the truncated OPE, 
and simultanously the contribution from excited and continuum states is smaller than thirty percents when summing up the hadrons. 
The threshold value $s_0$ is usually close to the outset of the first higher excited state with the same quantum number, 
then a certain vicinity can be expected, 
we determine it with considering the maximal stability of physical quantities once the Borel mass has been set down. 
Within the interval $M^2 = 1.0 \pm 0.1$ GeV$^2$ at the fixed threshold value $s_0 = 2.0 \pm 0.2$ GeV$^2$ 
which is slightly larger than the one used in \cite{Agaev:2017cfz,Agaev:2018sco} because we are discussing the $s\bar{s}$ current, 
we do the combined quadratic fit to both sides of Eqs.(\ref{eq:scalar-sr},\ref{eq:scalar-sr-1}), 
and obtain 
{\small
\beq
&&m_{f_0} = (985 \pm 122) \, \mathrm{MeV} \,, \non
&&m_{\sigma} = (439 \pm 304) \, \mathrm{MeV} \,, \non
&&\bar{f}^s_{f_0}(1 \, \mathrm{GeV}) = (358 \pm 4) \, \mathrm{MeV} \,, \non
&&\bar{f}^s_{\sigma}(1\, \mathrm{GeV})  \sim 0  \,.
\label{eq:f0sigma}
\eeq}
The $s-$flavor decay constant of $f_0$ agrees with the prediction $\bar{f}^s_{f_0} = (370 \pm 20) \, \mathrm{MeV}$ 
obtained under the assumption that $f_0(980)$ and $f_0(1500)$ are the lowest scalar states with $\bar{s}s$ assignment \cite{ChengNB}. 
The nearly zero $s-$flavor decay constant of $\sigma$ indicates that 
$f_0(980)$ is the lowest state in the channel with scalar isoscalar current $J^s$, 
standing by which we reevaluate the sum rules in Eqs.(\ref{eq:scalar-sr}, \ref{eq:scalar-sr-1}) by considering only the $f_0$ state, 
and obtain the same result for $m_{f_0}$ and $\bar{f}^s_{f_0}$ as listed in Eq.\ref{eq:f0sigma}.

\section{$\bar{B}_s \to f_0$ form factors from the LCSRs}\label{LCSRs}

The approach of LCSRs with B meson DAs was proposed to calculate the $B \to P,V$ form factors \cite{KhodjamirianST}, 
in this section we implement it to calculate the $\bar{B}_s \to f_0$ form factors. 

As we demonstrated in the last section that the scalar isoscalar current $\bar{s}s$ coupling to $\sigma$ is nearly zero, 
so it would be reasonable to consider $f_0(980)$ as the lowest scalar state in $B_s$ decays. 
This consideration is also supported by the fact that no evidence of $\sigma$ is observed in the $B$ decays so far \cite{AaijEMV}. 
Let's consider another correlation function 
{\small
\beq
&~& \Pi_{\nu}(p,q) = i \int d^4x \, e^{ip \cdot x} \, \langle 0 \vert T \{ J^s(x), J_\nu^\mathcal{I}(0) \} \vert \bar{B}_s(p+q) \rangle \,, \hspace{0.5cm}
\label{eq:corr-LCSRs}
\eeq}
with the weak current $J_\nu^\mathcal{I} = \bar{s} \Gamma_\nu^\mathcal{I} b$. 
The indicator $\mathcal{I} = A, T$ correspond to the gamma matrices 
$\Gamma_\nu^A = \gamma_\nu \gamma_5 $ and $\Gamma_\nu^T = \sigma_{\nu\mu}\gamma_5 q^\mu$, 
respectively\footnote{We use the convention $\sigma_{\mu\nu} = \frac{i}{2}(\gamma_\mu\gamma_\nu- \gamma_\nu\gamma_\mu)$.}. 
The heavy-to-light current is reduced to the light quark current after transiting to the heavy quark effective theory (HQET), 
and the correlation function is modified to
{\small
\beq
&~& \tilde{\Pi}_{\nu}(p,\tilde{q}) = 
i \int d^4x e^{ip \cdot x} \, \langle 0 \vert T \{ J^s(x), \tilde{J}_\nu^\mathcal{I}(0) \} \vert \bar{B}_{s,v}(p+\tilde{q}) \rangle \,.  \hspace{0.6cm}
\label{eq:corr-LCSRs-1}
\eeq}
We use the notations $p$ to denote the momentum carried by the scalar isoscalar current $J^s$, 
and $\tilde{q}= q-m_b v $ for the effective current $\tilde{J}^\mathcal{I}_\nu = \bar{s} \Gamma^\mathcal{I}_\nu h_v$. 
In the rest frame the effective $b$-quark field is defined by $h_v(x) = e^{i m_b v x} b(x)$ with the unit vector $v=(1,0,0,0)$, 
and $\vert \bar{B}_s(p+q)  \rangle= \vert \bar{B}_{s,v}(p+\tilde{q}) \rangle$ holds up to the $\mathcal{O}(1/m_b)$ accuracy.  
We would discuss in the intervals $\vert p^2 \vert, \vert \tilde{q}^2 \vert \gg \Lambda_{QCD}^2, (m_{B_s}-m_b)^2$ 
where the correlation function does not fluctuate violently and the OPE calculation works well. 

In the correlation function, the same flavour quark fields with small displacement can be contracted in the form of quark propagator, 
{\small
\beq
s_s(x,0,m_s) = \int \frac{d^4p}{(2\pi)^4} e^{-ipx} \int_0^1 du G_{\mu\nu}(ux) \non
\cdot \Big[ \frac{u x^\mu \gamma^\nu}{p^2-m_s^2} - \frac{(\psl+m_s)\sigma_{\mu\nu}}{2(p^2-m_s^2)^2} \Big] \,, 
\label{eq:quark-propagator}
\eeq}
in which the first term is the freedom part in the QCD limit, 
and the second term respects the soft one-gluon correction. 
Two-particle and three-particle DAs of $B_s$ meson are defined as \cite{GeyerFB}, 
{\small
\beq
&~&\langle 0 \vert \bar{s}_\alpha(x) h_{v \beta}(0) \vert \bar{B}_{s,v} \rangle \non
&=& -\frac{if_{B_s}m_{B_s}}{4} \int_0^\infty d \omega e^{-i \omega v \cdot x} \, 
\Big[(1+\vsl) \Big\{  \left[ \phi_+(\omega) + x^2 g_+(\omega) \right] \non
&~&- \frac{ \left[\phi_+(\omega) - \phi_-(\omega) + x^2 \left(g_+(\omega) - g_-(\omega) \right)\right]}{2v \cdot x} 
\xsl \Big\} \gamma_5 \Big]_{\beta\alpha} \,,
\label{eq:B-WFs-1}\\ 
\non
&~&\langle 0 \vert \bar{s}_\alpha(x) G_{\rho\delta}(ux) h_{v \beta}(0) \vert \bar{B}_{s,v} \rangle \non
&=& \frac{f_{B_s}m_{B_s}}{4}
\int_0^\infty d \omega \int_0^\infty d \zeta e^{-i (\omega+u\zeta)v \cdot x} \Big[ (1+\vsl) \non
&~& \cdot \Big\{(v_\rho \gamma_\delta - v_\delta \gamma_\rho) \, [\Psi_A(\omega,\zeta) - \Psi_V(\omega,\zeta)] 
- i \sigma_{\rho\delta} \Psi_V(\omega,\zeta)   \non
&~&- \left(\frac{x_\rho v_\delta - x_\delta v_\rho}{v \cdot x}\right) \, X_A(\omega,\zeta) \non
&~&+ \left(\frac{x_\rho \gamma_\delta - x_\delta \gamma_\rho}{v \cdot x}\right) \, \left[ Y_A(\omega,\zeta) +W(\omega,\zeta) \right]  \non
&~&+ i \epsilon_{\rho\delta\alpha\beta} \, \frac{x^\alpha v^\beta}{v\cdot x} \, \gamma_5 \, \tilde{X}_A(\omega,\zeta)
- i \epsilon_{\rho\delta\alpha\beta} \, \frac{x^\alpha \gamma^\beta}{v \cdot x} \, \gamma_5 \, \tilde{Y}_A(\omega, \zeta) \non
&~&- \left( \frac{x_\rho v_\delta - x_\delta v_\rho}{v \cdot x}  \right)\, \frac{\xsl}{v \cdot x} \, W(\omega, \zeta) \non
&~&+ \left( \frac{x_\rho \gamma_\delta - x_\delta \gamma_\rho}{v \cdot x} \right) \, \frac{\xsl}{v \cdot x}  \, Z(\omega,\zeta) \,
\Big\} \gamma_5 \Big]_{\beta\alpha} \,, 
\label{eq:B-WFs-2}
\eeq}
respectively. 
Two variables $\omega$ and $\zeta$ are introduced to represent the plus components of light quark and the gluon momentum, respectively. 
The path-ordered gauge factor is always underlied in the matrix element sandwiched between meson state and vacuum. 
Recently, the renormalization group equations (RGE) are resolved in the $N_C$ limit for three-particle DAs \cite{BraunLIQ}, 
and the models are suggested for the higher-twist $B$ meson DAs \cite{BenekeWJP},  
following which the power suppressed correction are supplemented to $B \to P, V, \gamma$ form factors 
\cite{Descotes-GenonBUD,BenekeWJP,GubernariWYI,LuCFC,ShenABS}.
The Lorentz definition in Eqs.(\ref{eq:B-WFs-1},\ref{eq:B-WFs-2}) should not be confused with the definite twist definition of LCDAs, 
we collect their relations, as well as the general model for the later one in appendix \ref{app:B-LCDAs}. 

The definition of $\bar{B}_s \to S $ transition form factors is quoted as \cite{WirbelJ,BauerBM}
{\small
\beq
&~& \langle S(p) \vert J_\nu^A(0) \vert \bar{B}_s(p+q) \rangle \,\non
&=& - i [ \mathcal{F}_+(q^2)p_\nu +  \mathcal{F}_-(q^2) q_\nu ]  \,\non
&=& -i  \mathcal{F}_1(q^2) \Big[ (2p+q)_\nu - \frac{m_{B_s}^2-m_S^2}{q^2} q_\nu \Big] \non
&~& - i  \mathcal{F}_0(q^2) \frac{m_{B_s}^2-m_S^2}{q^2} q_\nu \,,
\label{eq:ff-definition-1}\\
\non
&~& \langle S(p) \vert J_\nu^T(0) \vert \bar{B}_s(p+q) \rangle  \non 
&=& - \frac{ \mathcal{F}_T(q^2)}{m_{B_s}+m_S} \Big[q^2(2p+q)_\nu - (m_{B_s}^2-m_S^2)q_\nu \Big] \,. 
\label{eq:ff-definition-2}
\eeq }
The following relations are suggested for the form factors associated with axial-vector current in Eq.\ref{eq:ff-definition-1}, 
{\small
\beq
&~&\mathcal{F}_1(q^2) = \frac{ \mathcal{F}_+}{2}(q^2) \,, \\
&~&\mathcal{F}_0(q^2) \frac{(m_{B_s}^2 - m_S^2)}{q^2} \,\non
&=& \mathcal{F}_-(q^2) + \frac{ \mathcal{F}_+(q^2)}{2} \frac{(m_{B_s}^2-m_S^2-q^2)}{q^2} \,.
\label{eq:relation-f0-f+}
\eeq}

We calculate the correlation function in Eq.\ref{eq:corr-LCSRs-1} under the narrow width approximation, 
obtain the LCSRs result for $\bar{B}_s \to f_0(980)$ form factors,  
{\small
\beq
&~& m_{f_0}\bar{f}^s_{f_0} \, \mathcal{F}^{\bar{B}_s \to f_0}_+(q^2) \, e^{-m_{f_0}^2/M^2} \non
&=& f_{B_s} m_{B_s}^2 \, \Big\{\int_0^{\sigma_0} \, d \sigma \, e^{-s_q/M^2} \Big[ \phi_+(\sigma m_{B_s}) \non
&-& \frac{ \overline{\phi}_{\pm}(\sigma m_{B_s}) }{\bar{\sigma}m_{B_s}} 
- \frac{8 \bar{\sigma}^2 m_{B_s}^2  \, g_+ (\sigma m_{B_s}) }{(\bar{\sigma}^2 m_{B_s}^2 - q^2)^2} \non
&-& \frac{4 \bar{\sigma}\, g'_+(\sigma m_{B_s}) }{(\bar{\sigma}^2 m_{B_s}^2 - q^2)} \Big]  
+ \Delta  \mathcal{F}_+(q^2,s_0,M^2) \Big\} \,,
\label{eq:f+}\\
\non
&~& m_{f_0}\bar{f}^s_{f_0} \, \mathcal{F}^{\bar{B}_s \to f_0}_-(q^2) \, e^{-m_{f_0}^2/M^2} \non
&=& f_{B_s} m_{B_s}^2 \, \Big\{\int_0^{\sigma_0} \, d \sigma \, e^{-s_q/M^2} 
\Big[ -\frac{\sigma  \, \phi_+(\sigma m_{B_s})}{\bar{\sigma}} \non 
&-& \frac{\overline{\phi}_{\pm}(\sigma m_{B_s}) }{\bar{\sigma}m_{B_s}} 
+ \frac{8 \bar{\sigma} \sigma m_{B_s}^2 \, g_+ (\sigma m_{B_s})}{(\bar{\sigma}^2 m_{B_s}^2 - q^2)^2} \non
&+& \frac{4 \sigma\, g'_+(\sigma m_{B_s}) }{(\bar{\sigma}^2 m_{B_s}^2 - q^2)} \Big] 
+ \Delta  \mathcal{F}_-(q^2,s_0,M^2) \Big\} \,, 
\label{eq:f-}\\
\non
&~& \frac{2 m_{f_0} \bar{f}^s_{f_0}}{m_{B_s}+m_{f_0}} \, \mathcal{F}^{\bar{B}_s \to f_0}_{T}(q^2)  \, e^{-m_{f_0}^2/M^2} \non
&=& f_{B_s}m_{B_s}^2 \, \Big\{ \int_0^{\sigma_0} \, d \sigma \, e^{-s_q/M^2} \non 
&\cdot& \Big[ \frac{\phi_+(\sigma m_{B_s}) }{\bar{\sigma} m_{B_s}} 
- \frac{8\bar{\sigma}m_{B_s}\, g_+ (\sigma m_{B_s})}{(\bar{\sigma}^2 m_{B_s}^2 - q^2)^2}  \non
&-& \frac{4 \, g'_+(\sigma m_{B_s})}{m_{B_s} (\bar{\sigma}^2 m_{B_s}^2 - q^2)} \Big] \non
&+& \frac{1}{m_{B_s}^2-m_{f_0}^2-q^2} \, \Delta  \mathcal{F}_{T,q}(q^2,s_0,M^2)\Big\} \,. 
\label{eq:fT} 
\eeq}
The contributions proportional to light quark mass $m_s$ are not shown explicitly above. 
The dimensionless variable $\sigma \equiv \omega/m_{B_s}$ is the longitudinal momentum fraction of the light quark inside $\bar{B}_s$ meson, 
and the virtuality of internal quark is $s_q = m_{B_s}^2\sigma - (q^2\sigma - m_s^2)/\bar{\sigma}$. 
To process the calculation, we have defined an auxiliary distribution 
{\small
\beq
\overline{\phi}_\pm(\omega) \equiv \int_0^\omega d\tau [\phi_+(\tau) - \phi_-(\tau) ] \,,
\label{eq:phig+-}
\eeq }
with the boundary conditions $\overline{\phi}_\pm(0) = \overline{\phi}_\pm(\infty) = 0$. 
The derivation is $g'_+(\sigma m_{B_s}) = \left(d/d\sigma\right) g_+(\sigma m_{B_s})$. 
We give several comments in orders:    
\begin{itemize}
\item[(i)] With taking into account the quark mass effect, 
our results shown in Eqs.(\ref{eq:f+},\ref{eq:f-}) consist with the calculations of $B \to D_0^\ast$ form factors \cite{ShenMM}. 
\item[(ii)] $\mathcal{F}_{T,p}(q^2)$ and $\mathcal{F}_{T,q}(q^2)$ are obtained by matching the coefficients associated 
with different Lorentz structures (say, $p_\nu$ and $q_\nu$, respectively) with the matrix element sandwiched by the tenser current. 
In principle, they should be equal to each other when considering only the two-particle DAs\footnote{Because 
the coefficients of three-particle correlation in $\Delta \mathcal{F}_{T,p}$ may have the $1/q^2$ factor, 
we would take $\Delta \mathcal{F}_{T,q}$ for this part contribution in numerical analysis.}, so we use the unified notation $\mathcal{F}_{T}(q^2)$. 
\item[(iii)] In the heavy quark limit $\sigma \to 0$, 
we reproduce the relations 
%$ \mathcal{F}_T(q^2) = \frac{m_{B_s}}{m_{B_s}-m_S}  \mathcal{F}_-(q^2)$ in the heavy quark limit, and also the relations 
$\mathcal{F}_+(0) =  2m_{B_s}/(m_{B_s}+m_S) \, \mathcal{F}_T(0) $ and $\mathcal{F}_-(0) = 0$ at the full recoiled point. 
\end{itemize}

Multiplying both sides of Eq.\ref{eq:ff-definition-1} by $q^\nu$, 
we derive the matrix elements deduced by the pseudo-scalar current $J^P = i m_b  \bar{s} \gamma_5 b$, 
{\small
\beq
&& \langle S(p) \vert J^P(0) \vert B_s(p+q) \rangle = (m_{B_s}^2-m_S^2) \mathcal{F}_0(q^2) \,,
\label{eq:ff-definition-3}
\eeq}
which suggests another sum rules,  
{\small
\beq
&~& m_{f_0} \bar{f}^s_{f_0} \, (m_{B_s}^2-m_{f_0}^2) \, \mathcal{F}^{\bar{B}_s \to f_0}_0(q^2) \,  e^{-m_{f_0}^2/M^2} \, \non
&=& f_{B_s} m_{B_s}^2 m_b \,
\Big\{\int_0^{\sigma_0} \, d \sigma \, e^{-s_q/M^2} \non
&\cdot& \Big[ \frac{(\bar{\sigma}^2m_{B_s}^2-q^2 )\, \phi_+(\sigma m_{B_s}) }{2 \bar{\sigma}^2 m_{B_s}} 
- \frac{3\, \overline{\phi}_{\pm}(\sigma m_{B_s})}{\bar{\sigma}}  \non
&-& \left[2 \bar{\sigma} m_{f_0}^2 - 2 \sigma q^2 +(1-2\sigma) (m_{B_s}^2-m_{f_0}^2-q^2) \right] \, \non
&\cdot& \left(\frac{\bar{\sigma}m_{B_s} \, g_+(\sigma m_{B_s})}{(\bar{\sigma}^2m_{B_s}^2-q^2)^2} \, 
+ \frac{g'_+(\sigma m_{B_s})}{2m_{B_s}(\bar{\sigma}^2m_{B_s}^2-q^2)} \right)\Big]  \non
&+& \Delta \mathcal{F}_0(q^2,s_0,M^2) \Big\} \,. 
\label{eq:f0}
\eeq }
We remark that Eq.\ref{eq:ff-definition-3} is established on the heavy quark limit, 
so the new LCSRs in Eq.\ref{eq:f0} can be used to estimate how well does the heavy quark expansion work 
by comparing it with the original LCSRs in Eqs.(\ref{eq:f+},\ref{eq:f-}). 

The contributions from three-particle DAs of $B$ meson are arranged in an universal form 
{\small
\beq
&~& \Delta \mathcal{F}_i(q^2,s_0,M^2) \non 
&=& \int_0^{\sigma_0} \, d \sigma \, e^{-s_q/M^2}
\left(-I_{1,i}(\sigma) + \frac{I_{2,i}(\sigma)}{M^2} - \frac{I_{3,i}(\sigma)}{2M^4}\right) \non
&+& \frac{e^{-s_0/M^2}}{m_{B_s}^2} \, \Big\{ \eta(\sigma) \Big[ I_{2,i}(\sigma) 
- \frac{I_{3,i}(\sigma)}{2} \left(\frac{1}{M^2} + \frac{1}{m_{B_s}^2} \frac{d \eta}{d\sigma}\right) \non
&-& \frac{\eta}{2m_{B_s}^2}\frac{d I_{3,i}(\sigma)}{d\sigma} \Big]
\Big\} \Big\vert_{\sigma=\sigma_0} \,,
\label{eq:ff-3p}
\eeq}
where the dimensionless variable $\eta = (\bar{\sigma}^2 m_{B_s}^2)/(\bar{\sigma}^2 m_{B_s}^2-q^2+m^2)$ 
can be understood as the ratio between the minimal virtuality of the $b$ quark field 
and the maximal virtuality carried by the internal light quark.  
The integral over the three-particle DAs is written as 
{\small
\beq
I_{N,i}(\sigma) &=& \frac{1}{\bar{\sigma}^N} \int_0^{\sigma m_{B_s}} d\omega \int_{\sigma m_{B_s} -\omega}^\infty \frac{d \zeta}{\zeta} \non
&\cdot& \Big\{ C_{N,i}^{\Psi_A}(\sigma, u, q^2) \, \Psi_A(\omega, \zeta) \non
&+& C_{N,i}^{\Psi_V}(\sigma, u, q^2) \, \Psi_V(\omega, \zeta) \non
&+& C_{N,i}^{\overline{X}_A}(\sigma, u, q^2) \, \overline{X}_A(\omega, \zeta) \non
&+& C_{N,i}^{\overline{Y}_A}(\sigma, u, q^2) \, \left[ \overline{Y}_A(\omega, \zeta) + \overline{W}(\omega, \zeta) \right] \non
&+& C_{N,i}^{\overline{\widetilde{X}}_A}(\sigma, u, q^2) \,\overline{\widetilde{X}}_A(\omega, \zeta) \non
&+& C_{N,i}^{\overline{\widetilde{Y}}_A}(\sigma, u, q^2) \, \overline{\widetilde{Y}}_A(\omega, \zeta) \non
&+& C_{N,i}^{\overset{=}{W}}(\sigma, u, q^2) \, \overset{=}{W}(\omega, \zeta) \non
&+& C_{N,i}^{\overset{=}{Z}}(\sigma, u, q^2) \, \overset{=}{Z}(\omega, \zeta)
 \Big\}  \Big\vert_{u=(\sigma m_{B_s} - \omega)/\zeta} \,,
\label{eq:int-3p}
\eeq}
here another two auxiliary distributions are introduced for 
$\overline{X}_A \,, \overline{Y}_A \,, \overline{W} \,, \overline{\widetilde{X}}_A \,, \overline{\widetilde{Y}}_A$ and $\overset{=}{W} \,, \overset{=}{Z}$,
{\small
\beq
\overline{f} \equiv \int_0^\omega d \tau \, f(\tau, \zeta) \,, \,\,\,\,\,\,
\overset{=}{f} \equiv \int_0^\omega d \tau \int_0^\zeta d \tau' \, f(\tau, \tau') \,. 
\eeq}
The lower indicator $N=1,2,3$ stands for the power of Borel mass $M^{-2(N-1)}$ premultiplied with the integrals, 
the coefficients $C_{N,i}$ associated to each three-particle DA are presented in appendix \ref{app:coeff-3p}.

\begin{figure}
\centering{
\resizebox{0.42\textwidth}{!}{%
\includegraphics{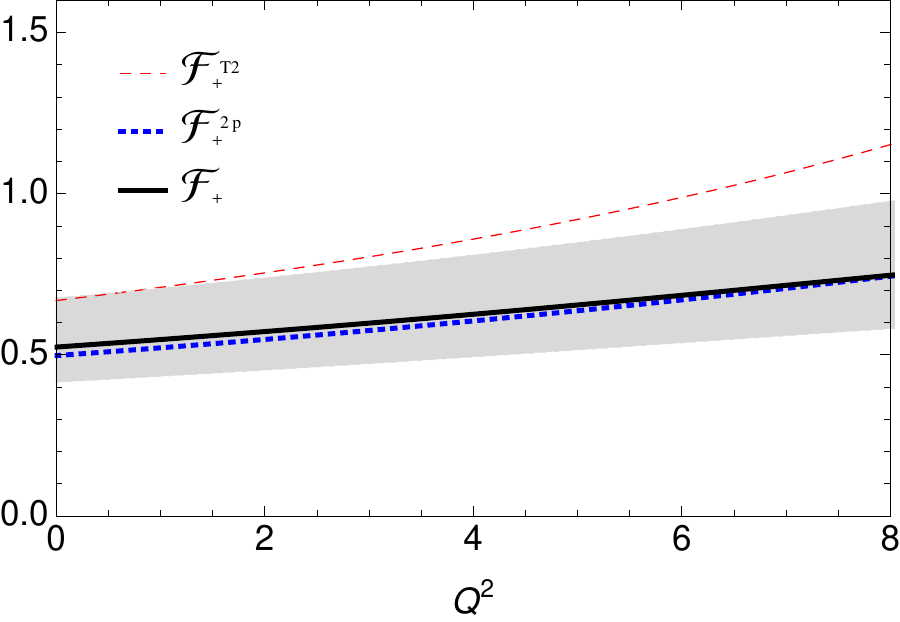} }
\resizebox{0.42\textwidth}{!}{%
\includegraphics{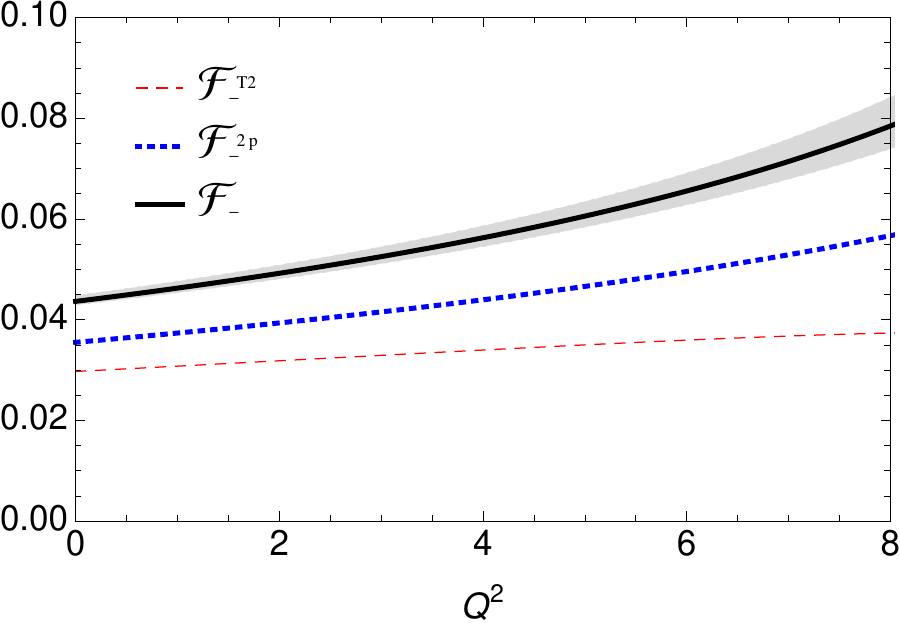}}
\resizebox{0.42\textwidth}{!}{
\includegraphics{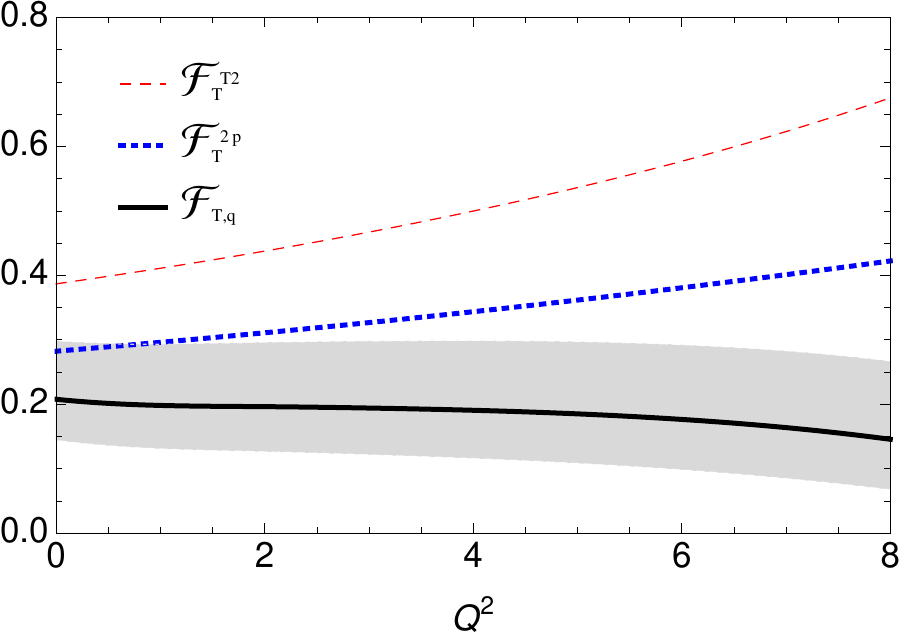}}
\resizebox{0.42\textwidth}{!}{%
\includegraphics{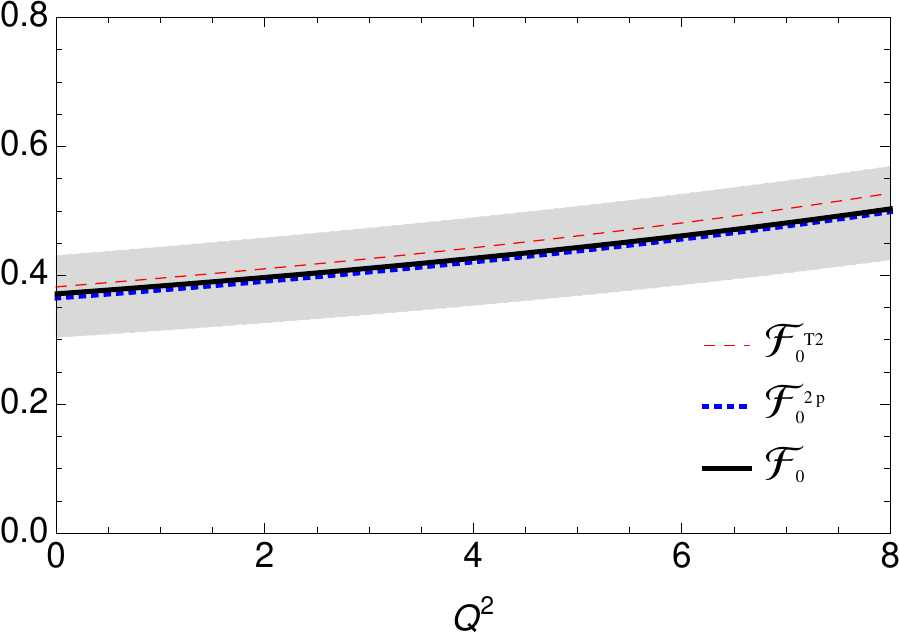}}
}
%\vspace{5cm}       % Give the correct figure height in cm
\caption{LCSRs predictions for $\bar{B}_s \to f_0$ form factors with $Q^2 = q^2$, 
where the red-dashed and blue-dotted curves represent the contributions with considering only the leading twist and also the high twists  
two-particle DAs of $\bar{B}_s$ meson, respectively. 
The black curves indicate the form factors obtained with including both two- and three-particle DAs.}
\label{fig:1}       % Give a unique label
\end{figure}

We take $m_b(m_b) = 4.2$ GeV for $b$ quark mass in the typical $\overline{\mathrm{MS}}$ scheme, 
and use $f_{B_s} = 0.242$ GeV obtained from lattice QCD \cite{BazavovAA} and two-point QCD sum rules \cite{GelhausenWIA}. 
The inverse moment of $\bar{B}_s$ DAs is chosen in the interval $\lambda_{\bar{B}_s} = 450 \pm 50$ MeV, 
a little bit smaller then the conventional value $500 \pm 50$ MeV, 
by considering the possible next-to-leading-order radiative correction effects\footnote{Where 
$\lambda_B = 358^{+38}_{-30} (343^{+22}_{-20})$ MeV is obtained by comparing the $B \to \pi (\rho)$ form factor from LCSRs with pion \cite{KhodjamirianUB} (rho \cite{StraubICA}) and $B$ meson DAs.} 
\cite{WangVGV,GaoLTA}. 
The same ballpark of the LCSRs parameters, say, $M^2 = 1.0 \pm 0.1$ GeV$^2$ and $s_0 = 2.0 \pm 0.2 $ GeV$^2$, 
are used here as in the two-point sum rules in the last section, 
with which the OPE convergence is automatically manifested by the relative small three-partical DAs contribution 
$\mathcal{F}^{3p}(Q^2)/\mathcal{F}^{2p}(Q^2) \lesssim 30\%$ for $Q^2 \in [0, 5]$ GeV$^2$. 
We plot in figure \ref{fig:1} for the $\bar{B}_s \to f_0$ form factors obtained under the narrow width approximation,  
where the lightgray shadows reveal the total uncertainty came from the inverse moment $\lambda_{\bar{B}_s}$ 
and the LCSRs parameters.  
%It is obvious to check the relations between the form factors in the heavy quark limit at the full recoiled energy point. 
The high twist $B$ meson DAs with two-particle configuration ($g_+$) give about $25\%$ decrease to $\mathcal{F}_{+}$ and $\mathcal{F}_{T}$, 
about $20\%$ increase to $\mathcal{F}_{-}$, 
the tiny change of $\mathcal{F}_0$ from $g_+$ can be understood by the interplay 
between the negative correction for $\mathcal{F}_{+}$ and the positive correction for $\mathcal{F}_{-}$. 
The three-particle $\bar{B}_s$ DAs bring another $25\%$ decrease and increase to $\mathcal{F}_{T}$ and $\mathcal{F}_{-}$, respectively, 
while its corrections to $\mathcal{F}_+$ and $\mathcal{F}_0$ are tiny, 
indicating that the heavy quark limit is broken with considering the three-particle DAs correction. 
We compare our result for $\bar{B}_s \to f_0$ form factors with other methods in table \ref{tab1}. 

\begin{table}
\caption{$\bar{B}_s \to f_0(980)$ form factors at $q^2=0$ ($q^2 = 4m_\pi^2$ for \cite{ElBennichXY}) predicted in different methods.}
\label{tab1}       % Give a unique label
\begin{center}
\begin{tabular}{c|ccc}
\toprule[1pt]%\hline\noalign{\smallskip}
$\mathrm{Methods}$ & $\mathcal{F}_+$  & $\mathcal{F}_-$ & $\mathcal{F}_T$   \\
\midrule[0.5pt]%\noalign{\smallskip}\hline\noalign{\smallskip}
$\mathrm{PQCD}$ \cite{LiTK} & $0.70$ & $\equiv 0$ & $0.40$ \non
$\mathrm{CLFD}$ \cite{ElBennichXY} & $0.80$ & $\equiv 0$ & $-$  \non
$\mathrm{CQM}$ \cite{IssadykovIBA}& $0.254$ & $-$ & $0.285$  \non
\midrule[0.4pt]
$\mathrm{QCDSRs} $ \cite{GhahramanyZZ} & $0.12$ & $-0.07$ & $-0.08$ \non
$\mathrm{LCSRs} $ \cite{ColangeloBG} & $0.37$ & $\equiv 0 $ & $0.228$ \non
$\mathrm{LCSRs-chiral} $ \cite{SunNV} & $0.44$ & $-0.44$ & $0.58$ \non
$\mathrm{LCSRs} $ \cite{WangVRA} & $0.90$ & $0.14$ & $0.60$ \non
\midrule[0.4pt]
$\mathrm{this \, work}$ & $0.52$ & $0.04$ & $0.21$ \non
\bottomrule[1pt]%\noalign{\smallskip}\hline
\end{tabular}
\end{center}
%\vspace*{0cm}  % with the correct table height
\end{table}

\section{The width effect and the $\bar{B}_s \to KK$ form factors}\label{width-effect}

To investigate the width effect of intermediate states, 
let's look back to the dispersion relation of correlation function in Eq.\ref{eq:corr-LCSRs},  
{\small
\beq
\Pi_\nu(p,q) = \frac{1}{\pi} \int_0^\infty ds \frac{\mathbf{Im} \Pi_\nu(s, q^2) }{s-p^2-i\epsilon} \,. 
\label{eq:DR}
\eeq}
The imaginary part in the numerator, corresponding to the physical regions ($q^2 > 0$), 
can be obtained by interpolating a complete set of intermediate states between two local currents in the correlation function, 
{\small
\beq
&&2 \, \mathbf{Im} \Pi_\nu(s, q^2) = \sum_{n} \, \int d \tau_n \, (2\pi)^4 \delta(s - p_n^2) \, \non
&&\cdot \langle 0 \vert J^s \vert S_n(p) \rangle  \, 
\langle S_n(p) \vert J_\nu^\mathcal{I}(q^2) \vert \bar{B}_s(q+p_n) \rangle  \,,
\label{eq:unitaty-relation}
\eeq }
where $p$ and $d\tau_n$ denote the momentum and phase space volume of each state $\vert S_n \rangle$, respectively. 
In the narrow width approximation, $\vert S_n \rangle$ are single mesons and the dispersion relation reduces to
{\small
\beq
\Pi_\nu(p^2,q^2) &=&  \frac{m_{f_0} \bar{f}^s_{f_0} \, 
\langle f_0(p) \vert J^\mathcal{I}_\nu(q) \vert \bar{B}_s(p+q) \rangle}{m_{f_0}^2 - p^2 -i \epsilon} \, \non
&+& \frac{1}{\pi} \int_{s_0}^\infty \, ds \, \frac{\mathbf{Im} \Pi_\nu(s,q^2)}{s-p^2-i\epsilon} \,. 
\label{eq:DR-1}
\eeq}
In Eq.\ref{eq:DR-1}, only the contribution from ground state is singled out while the rest parts are retained in the integral,  
this is exactly what we did in the last section. 

A straightforward way to consider the width effect is to substitute the interpolation of single mesons by stable multi-meson states, 
such as the $KK, \, \pi\pi, \, \eta\eta$ and their continuum states \cite{ChengSMJ},  
{\small
\beq
&~&\Pi_\nu(k^2,q^2) \, = \frac{1}{\pi} \int_{4m_K^2}^{s_0^{2 K}} \, ds \, \int d\tau_{2K} \,  \non
&\cdot& \frac{ \langle 0 \vert J^{s} \vert KK_{I=0} \rangle \, 
\langle KK_{I=0} \vert J^\mathcal{I}_\nu(q) \vert \bar{B}_s(k+q) \rangle}{s - k^2 -i \epsilon} \, + \cdots \,. 
\label{eq:DR-2}
\eeq}
Here we only write out explicitly the term contributed from $KK$ state, 
the ellipsis denotes the contributions from $\pi\pi$, $4\pi$, $\eta\eta$, $\eta\eta^\prime$ and their excited states. 
Because it is much more harder to generate a $\bar{s}s$ pair than a $\bar{n}n$ pair from vacuum, 
the contribution from $\eta\eta$, $\eta\eta^\prime$ channels can be expected to be small. 
The contributions from $\pi\pi$ and $4\pi$ channels are also small since they are produced via $KK \to \pi\pi, 4\pi$ rescattering, 
which exceeds the scope we are discussing here. 
%We do not take in to account for the radiative and leptonic channels $\gamma\gamma$ and $e^+e^-$ contributed via the annihilation mechanism. 
In the follow calculation we consider $KK$ state as the ground state which gives dominant contribution to the correlation function, 
while the contributions from high excited multi-meson states are suppressed by the Borel exponent $e^{-s/M^2}$ (with the Borel mass $M^2 \lesssim 1.1$ GeV$^2$). 

\subsection{Formalism}\label{sub:formulas}

The scalar isoscalar kaon form factor is defined as \cite{DonoghueXH}
{\small
\beq
\big\langle K^\rho(k_1) K^\sigma(k_2)_{I=0}  \, \big\vert \, J^s \, \big\vert \, 0 \big\rangle 
= \frac{\varGamma_K^s(k^2)}{m_s} \, \delta^{\rho\sigma}\,.
%= \sqrt{2} \frac{\varGamma_K^s(k^2)}{m_s} \,.
\label{eq:ff-S}
\eeq}
There is no experiment measurement for $\Gamma_K^s$ as yet\footnote{For the scalar isoscalar pion form factor $\Gamma_\pi$ defined with $\bar{n}n$ source current, 
the $N$-subtracted omn\'es representation gives a good description for the data up to $k^2 = 1.5^2$ GeV$^2$ 
with considering the generalized Watson theorem of the $\pi\pi \to \pi\pi$ phase 
\cite{YndurainCM,PelaezVS,KaminskiQE,GarciaMartinCN,KangJAA,YuanFOA}. }, 
so our calculation relies on the theoretical input.
%For the kaon form factor in our interesting, strictly speaking, 
%the scalar isoscalar $\pi\pi$ system is coupled to the $\bar{K}K$ channel via the $f_0(980)$ resonance, 
%and a coupled-channel description becomes inevitable even for energies around s = 1 GeV2. 
In the chiral perturbative theory (CHPT), 
$\varGamma_K^s$ has been derived to next-to-leading-order \cite{MeissnerBC} with supplementing by the unitarity constraint \cite{LahdeWR}, 
however, the unitarized CHPT works only at low energies, say, $s < 1.1$ GeV$^2$ \cite{DoringWKA,MeissnerHYA}. 
In order to include the high energy behaviour, 
one is forced to employ a model \cite{Hanhart:2012wi,RopertzSTK} 
and/or to adopt the perturbative QCD approximation \cite{ChengKHI}. 

From the view of hadron, $\varGamma_K^s$ is relevant to the $T$ matrix elements of 
$\pi\pi \to KK$ and $KK \to KK$ scatterings \cite{Hanhart:2012wi,RopertzSTK} via
\beq
\Gamma_K^s (k^2) = M_K(k^2) + T_{Ki} G_{ii} M_{i} \,,
\label{eq:Gamma-T}
\eeq 
where $M_i$, in principle, is an analytic term describing the transition from the scalar isoscalar source to the channel $i$ 
($i=1$ represents for $\pi\pi$ and $i=2$ for $KK$), 
and $G_{ii}$ is the free propagation of the particles in channel $i$. 
The $\pi\pi$ channel does not contribute to the $\bar{s}\Gamma_\nu^{\mathcal{I}}b$ transition at Born level, 
which is one of the reasons we can obtain the approximate dispersion relation in Eq.\ref{eq:DR-2}, 
with retaining only the $KK$ channel. 
To maintain the self-consistency of the intrepolation, 
we take only the $KK$ channel in Eq.\ref{eq:Gamma-T} too, 
and the transition from source current $J^s$ to $KK$ channel happens as $G_{KK} M_{K} = 1$ with subsequently the free propagating . 
In this way, a simple relation is obtained as 
\beq
\Gamma_K^s (k^2) = M_K(0) + T_{KK}(k^2) \,,
\label{eq:Gamma-T-1}
\eeq
with the normalisation $M_K(0) = m_K^2 - m_\pi^2/2$. 

%%-----------------------------------------------------------------------
\begin{figure}
\resizebox{0.45\textwidth}{!}{
\includegraphics{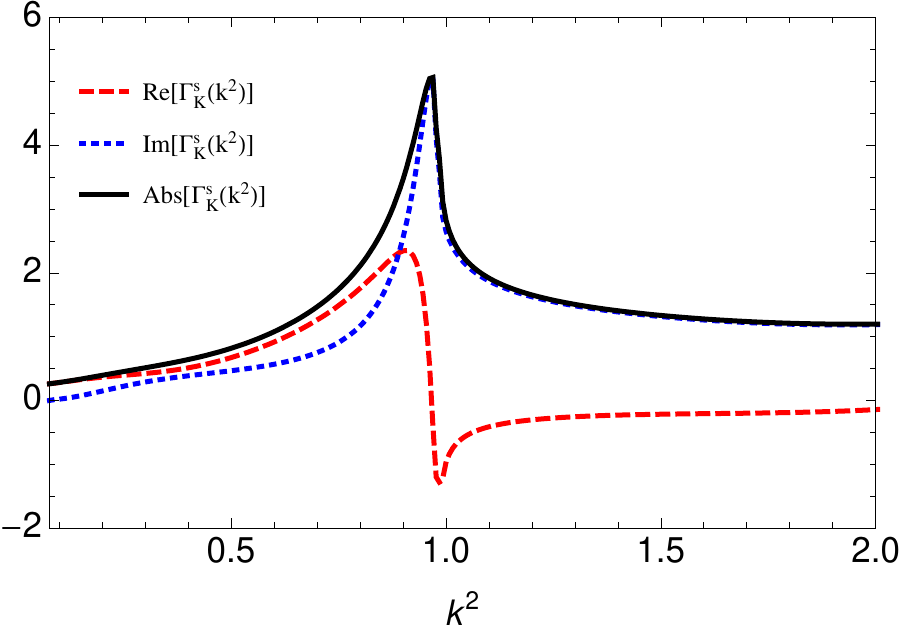}
}
\caption{The moduli of $\varGamma_K^s$ obtained from $I=J=0$ hadronic $KK \to KK$ scattering.}
\label{fig:2}       
\end{figure}
%%-----------------------------------------------------------------------

We can read from Eqs.(\ref{eq:Gamma-T},\ref{eq:Gamma-T-1}) that the phase of $s$ flavoured kaon form factor 
is only determined by the $KK \to KK$ scattering amplitude, 
whose expression with definite isospin ($I$) and partial wave ($J$) is parameterised as 
\beq
T^I_J(k^2) &=& \frac{1}{2 i \beta_K(k^2)} \Big[ \eta^I_J(k^2) \, e^{2 i \delta^I_J(k^2)} -1 \Big] \non
&=& \vert T^I_J(k^2) \vert \, e^{i \phi^I_J(k^2)} \,.
\label{eq:T-matrix}
\eeq
In the above equation, $\beta_K(k^2) = \sqrt{1-4m_K^2/k^2}$ is the phase space of $KK$ system, 
$\eta^I_J$ and $\delta^I_J$ are the inelasticity and phase shift, respectively, 
and $\phi^I_J$ is the phase. 
We would use the result of $T^0_0(k^2)$ obtained from the amplitude analysis \cite{DaiZTA,DaiUAO} 
as input for the scalar form factor\footnote{We thanks Ling-yun Dai for sharing us with the original result of their global fit analysis.}, 
which, as they claimed, can be extrapolated to a high energy $\sim 5$ GeV$^2$. 
We show in figure \ref{fig:2} for the result in the energy regions $k^2 \in [4m_\pi^2, 2.0]$ GeV$^2$, 
these curves consist with the result obtained from CHPT in the low energy regions $k^2 < 1.1$ GeV$^2$ \cite{DoringWKA}, 
and also consist with the fully $K$-matrix description in the high energies \cite{RopertzSTK}. 
The amplitude analysis result we adopted here considered all the measured data, the $\pi\pi-K\bar{K}$ final state interaction, 
the mass difference between the charged and neutral kaon, and the low energy Roy equation \cite{BuettikerPP}. 
In the amplitude analysis demonstrated by coupled-channel treatment and combined fitting, 
the source current with $s\bar{s}$ configuration is overwhelming coupled to $KK$ channel through $f_0$, 
while the coupling to $\pi\pi$ is tiny.

The remaining matrix element in Eq.\ref{eq:DR-2} are defined in terms of $\bar{B}_s \to KK$ transition form factors \cite{FallerDWA}, 
for the axial-vector current $j_\nu^A$ we have
{\small
\beq
&~&-i \langle K^+(k_1)K^-(k_2) \vert \bar{s} \gamma_\nu \gamma_5 b \vert \bar{B}_s(q+k) \rangle \non
&=& F_t \frac{q_\nu}{\sqrt{q^2}} 
+ F_0 \frac{2\sqrt{q^2}}{\sqrt{\lambda_B}} \Big(k_\nu - \frac{k \cdot q}{q^2}q_\nu \Big) \non 
&+&  \frac{F_\parallel}{\sqrt{k^2}} \Big( \bar{k}_\nu - \frac{4(q \cdot k) (q \cdot \bar{k})}{\lambda_B} k_\nu + 
\frac{4k^2(q \cdot \bar{k})}{\lambda_B} q_\nu \Big) \,.   
\label{eq:B2dipi-ff-1} 
\eeq}
The kinematics of $\bar{B}_s$ decay to $KK$ state are described by three independent variables: $k^2$, $q^2$ and $\theta_K$, 
denoting the invariant mass of $KK$ system, the squared momentum transfer in the weak decay 
and the angle between the 3 momentum of $K^-(k_2)$ and $\bar{B}_s$ meson in the $K^+K^-$ rest frame, respectively.  
The dot products are 
{\small
\beq
&&q \cdot k = \frac{1}{2}(m_{B_s}^2-k^2-q^2) \,,  \non
&&q \cdot \bar{k} = \frac{\sqrt{\lambda_B}}{2} \beta_K(k^2) \cos \theta_K \,,
\eeq}
with  the kinematic $\mathrm{K\ddot{a}ll\acute{e}n}$ function 
$\lambda_B \equiv \lambda(m_{B_s}^2,q^2,k^2) = m_{B_s}^4 + k^4 +q^4 -2(m_{B_s}^2k^2+m_{B_s}^2q^2+k^2q^2)$. 
By the way, the matrix element in Eq.\ref{eq:B2dipi-ff-1} can also be defined by the helicity amplitudes, 
{\small
\beq
H_\lambda = \langle K^+(k_1) K^-(k_2) \vert \bar{s} \gamma_\nu \gamma_5 b \vert \bar{B}_s(q+k) \rangle \,.
\label{eq:B2dipi-ff-helicity}
\eeq }
The helicity definition provides a possibility to study the contributions from different partial waves, 
because $H_\lambda$, with $\lambda = t, 0, + ,-$, can be expanded in terms of the associated Legendre polynomials. 
To study the partial waves contributions within the convenient definition in orthogonal Lorentz structures, 
we translate the partial wave expansion from the helicity amplitudes $H_\lambda$ to the form factors $F_{i}$, 
{\small
\beq
&&F_{0,t}(q^2,k^2,q \cdot \bar{k}) = \sum_{l=0}^{\infty} \sqrt{2l+1} \, F_{0,t}^{(l)}(q^2,k^2) P_l^{(0)}(\cos \theta_\pi) \,, \non
&&F_{\parallel}(q^2,k^2,q \cdot \bar{k}) = \sum_{l=1}^{\infty} \sqrt{2l+1} \, F_{\parallel}^{(l)}(q^2,k^2) \frac{P_l^{(1)}(\cos \theta_\pi)}{\sin \theta_\pi} \,. 
\label{eq:partial-expansion}
\eeq}

Considering the decomposition of isoscalar $KK$ state,
{\small
\beq
%\vert \pi\pi_{I=0} \rangle = \frac{1}{\sqrt{2}} \left( \frac{\sqrt{2}}{\sqrt{3}} \vert \pi^+\pi^- \rangle 
%+ \frac{1}{\sqrt{3}} \vert \pi^0\pi^0 \rangle \right) \,,
\vert KK_{I=0} \rangle =  \frac{1}{\sqrt{2}}  \vert K^+K^- \rangle + \frac{1}{\sqrt{2}} \vert K^0 \bar{K}^0 \rangle \,,
\eeq}
substituting Eq.\ref{eq:B2dipi-ff-1} and Eq.\ref{eq:partial-expansion} into Eq.\ref{eq:DR-2}, 
we obtain the $S-$wave contribution to the imaginary part with interpolating scalar isoscalar $KK$ state,   
{\small
\beq
&~&\int d\tau_{2K} \, 
\langle 0 \vert J^{s} \vert KK_{I=0} \rangle \, 
\langle KK_{I=0} \vert J^\mathcal{A}_\nu(q) \vert \bar{B}_s(k+q) \rangle \, \non
&=& \int d\tau_{2K} \, 2 \, \frac{\varGamma^{s \ast}_K(k^2)}{m_s} \, 
\langle K^+(k_1) K^-(k_2) \vert J_\nu^A(q) \vert \bar{B}_s(p+k) \rangle \non
&=& 2i \frac{\beta_K(s)}{8\pi} \frac{\varGamma^{s \ast}_K(s)}{m_s} \Big[ 
F_0^{(l=0)}(q^2,s) \frac{2\sqrt{q^2}}{\sqrt{\lambda_B}} k_\nu  \non
&+& \Big(\frac{F_t^{(l=0)}(q^2,k^2)}{\sqrt{q^2}} 
- F_0^{(l=0)}(q^2,s) \frac{2\sqrt{q^2}}{\sqrt{\lambda_B}}\frac{k \cdot q}{q^2}\Big)q_\nu \Big] \,.
\label{eq:numerator-1}
\eeq}
%To obtain this, we have acquiesced in the isospin relation 
%$\langle K^+K^- \vert J_\nu^\mathcal{A} \vert B_s \rangle = \langle \K^0\\bar{K}^0 \vert J_\nu^\mathcal{A} \vert B_s \rangle$. 
In fact, the phase space $d \tau_{2K}$ plays as a $S-$wave projector 
for the timelike-helicity form factors $F_{t,0}$, 
which means that only the $S-$wave component $F_{t,0}^{(l=0)}$ survives after integrating over the angle $\theta_K$. 
For the form factor $F_\parallel$, 
the role of $S-$wave projector vanishes and the contribution starts from $D$-wave component ($l=2n, n=1,2,3 \cdots$), 
which part is expected tiny in the $\bar{B}_s \to KK $ transtion and would not be discussed in this paper.

We take the global duality to eliminate the contributions beyond $KK$ state with the threshold $s_0^{2K}$, 
{\small\beq
&~&\frac{1}{\pi}\int_{s_0^{2K}}^\infty ds \, e^{-s/M^2} \, \mathbf{Im} \, \Pi_\nu (s,q^2) \non
&=& \frac{1}{\pi}\int_{s_0^{2K}}^\infty ds \, e^{-s_q/M^2} \, \mathbf{Im} \, \Pi^{\mathrm{OPE}}_\nu (s_q,q^2) \,, 
\label{eq:duality-1}
\eeq}
and arrive at the LCSRs result for $S-$wave $\bar{B}_s \to KK$ transition,  
{\small
\beq
&~&\int_{4m_K^2}^{s_0^{2K}} ds \, e^{-s/M^2} \frac{\beta_K(s)}{4 \pi^2} \, \frac{\varGamma^{s \ast}_K(s )}{m_s} \,
F_0^{(l=0)}(q^2,s) \frac{\sqrt{q^2}}{\sqrt{\lambda_B}}  \non
&=& f_{B_s}m_{B_s}^2 \Big\{ \int_0^{\sigma_0^{2K}} d\sigma \, e^{-s_q/M^2} \Big[ \phi_+(\sigma m_{B_s}) \non 
&-& \frac{\overline{\phi}_{\pm}(\sigma m_{B_s}) }{\bar{\sigma}m_{B_s}} 
- \frac{8 \bar{\sigma}^2 m_{B_s}^2 \, g_+ (\sigma m_{B_s})}{(\bar{\sigma}^2 m_{B_s}^2 - q^2)^2}  \non
&-& \frac{4 \bar{\sigma} \, g'_+(\sigma m_{B_s})}{(\bar{\sigma}^2 m_{B_s}^2 - q^2)}  \Big]  
+ \Delta  \mathcal{F}_+(q^2,s_0^{2K},M^2) \Big\} \,,
\label{eq:F0l=0}
\eeq
\beq
&~&\int_{4m_K^2}^{s_0^{2K}} ds \, e^{-s/M^2}  \frac{\beta_\pi(s)}{8 \pi^2} \, \frac{\varGamma^{s \ast}_K(s)}{m_s} \non
&\cdot&\Big(\frac{F_t^{(l=0)}(q^2,k^2)}{\sqrt{q^2}} - F_0^{(l=0)}(q^2,s) \frac{2\sqrt{q^2}}{\sqrt{\lambda_B}}\frac{k \cdot q}{q^2} \Big) \non
&=& f_{B_s}m_{B_s}^2 \Big\{ \int_0^{\sigma_0^{2K}} d\sigma \, e^{-s_q/M^2} \Big[ -\frac{\sigma }{\bar{\sigma}} \, \phi_+(\sigma m_{B_s}) \non
&-& \frac{\overline{\phi}_{\pm}(\sigma m_{B_s}) }{\bar{\sigma}m_{B_s}} 
+ \frac{8 \bar{\sigma} \sigma m_{B_s}^2  \, g_+ (\sigma m_{B_s})}{(\bar{\sigma}^2 m_{B_s}^2 - q^2)^2} \non
&+& \frac{4 \sigma \, g'_+(\sigma m_{B_s})}{(\bar{\sigma}^2 m_{B_s}^2 - q^2)}  \Big] 
+ \Delta  \mathcal{F}_-(q^2,s_0^{2K},M^2) \Big\} \,, 
\label{eq:Ft0l=0}
\eeq}

Multiplying both sides of Eq.\ref{eq:numerator-1} by $q^\nu$, 
%the phase space integral of the $B \to \pi\pi$ matrix element with pseudo-scalar current is modified to
%\beq
%&~&\int d\tau_{2\pi} \, \frac{\varGamma^\ast_\pi(k^2)}{2\hat{m}} \, 
%\langle \pi^+(k_1) \pi^-(k_2) \vert J^P \vert B^-(p+k) \rangle
%= i \frac{\varGamma^\ast_\pi(s)}{\hat{m}} \frac{\beta_\pi(s)}{16\pi} \sqrt{q^2} F_t^{(l=0)}(q^2,s) \,,   
%\label{eq:numerator-2}
%\eeq
we obtain another independent LCSRs for the timelike-helicity form factor $F_t^{(l=0)}(q^2,s)$, 
{\small
\beq
&~&\int_{4m_K^2}^{s_0^{2K}} ds \, e^{-s/M^2} \, \frac{\beta_\pi(s)}{8 \pi^2} \, \frac{\varGamma^{s \ast}_K(s )}{m_s} \, \sqrt{q^2} \,F_t^{(l=0)}(q^2,s)   \non
&=&f_{B_s}m_{B_s}^2 m_b \Big\{ \int_0^{\sigma_0^{2K}} d\sigma \, e^{-s_q/M^2} \,\non
&\cdot& \Big[
\frac{(\bar{\sigma}^2m_{B_s}^2 - q^2) \, \phi_+(\sigma m_{B_s})}{2\bar{\sigma}^2m_{B_s}} \,
-\frac{3 \, \overline{\phi}_\pm(\sigma m_{B_s}) }{\bar{\sigma}} \non
&-& \left[2 \bar{\sigma} m_{f_0}^2 - 2 \sigma q^2 +(1-2\sigma) (m_{B_s}^2-m_{f_0}^2-q^2) \right] \, \non
&\cdot& \left(\frac{\bar{\sigma}m_{B_s} \, g_+(\sigma m_{B_s})}{(\bar{\sigma}^2m_{B_s}^2-q^2)^2} \, 
+ \frac{g'_+(\sigma m_{B_s}) }{2m_{B_s}(\bar{\sigma}^2m_{B_s}^2-q^2)} \, \right)\Big]  \, \non
&+& \Delta \mathcal{F}_0(q^2,s_0^{2K},M^2) \Big\} \,.
\label{eq:Ftl=0}
\eeq}

\subsection{Models}\label{sub:models}

Eqs.(\ref{eq:F0l=0},\ref{eq:Ft0l=0},\ref{eq:Ftl=0}) are the main results in this section. 
Due to the convoluted integral, we can not solve out the form factors $F_{t/0}^{(l=0)}$ in terms of the $B$ meson DAs. 
To propel the calculation, one way we can try is to introduce the parameterisation of $S-$wave $\bar{B}_s \to KK$ form factors, 
and the first candidate coming into our mind is the single resonance ($f_0$) model 
in the generalized Breit-Wigner formula\footnote{We drop the $\sigma$ with the same reasons as described in section \ref{LCSRs}, 
which, phenomenologically, is further supported by the fact that no any signal is found for $KK$ coupling to $\sigma$ \cite{TanabashiOCA}.}. 
{\small
\beq
&&F_0^{(l=0)}(s,q^2) \frac{\sqrt{q^2}}{\sqrt{\lambda_B}} = 
\frac{1}{\sqrt{2} } \frac{g_{f_0 KK}\, \mathcal{F}^{\bar{B}_s \to f_0}_+(q^2) }{m_{f_0}^2 - s - i \sqrt{s} \, \Gamma_{f_0}(s)} \, e^{i\phi_{f_0}(s,q^2)}\,, 
\label{eq:ff-b2dipi-model-1} \\
&&\frac{1}{\sqrt{2}} \Big(\frac{F_t^{(l=0)}(s,q^2)}{\sqrt{q^2}} - F_0^{(l=0)}(s,q^2) \frac{\sqrt{q^2}}{\sqrt{\lambda_B}} \frac{m_{B_s}^2-s-q^2}{q^2} \Big) \non
&&= \frac{g_{f_0 KK} \, \mathcal{F}^{\bar{B}_s \to f_0}_-(q^2)}{m_{f_0}^2 - s - i \sqrt{s} \, \Gamma_{f_0}(s)}  \, e^{i\phi_{f_0}(s,q^2)} \,, 
\label{eq:ff-b2dipi-model-2} \\
&&\frac{1 }{\sqrt{2}} F_t^{(l=0)}(s,q^2) \sqrt{q^2} \non
&&= \frac{g_{f_0 KK} \, \mathcal{F}^{\bar{B}_s \to f_0}_0(q^2) (m_{B_s}^2- m_{f_0}^2) }{m_{f_0}^2 - s - i \sqrt{s} \, \Gamma_{f_0}(s)}  
\, e^{i\phi_{f_0}(s,q^2)}\,, \non
\label{eq:ff-b2dipi-model-3} 
\eeq}
The strong coupling $g_{f_0\pi\pi}$ is normalized as
{\small
\beq
\langle K^+(k_1)K^-(k_2) \vert f_0^s(k_1+k_2) \rangle = g_{f_0 K^+K^-} =\frac{ g_{f_0 KK}}{ \sqrt{2}} \,.
\label{eq:gf0pipi}
\eeq}
An underlying condition implied in Eqs.(\ref{eq:F0l=0},\ref{eq:Ft0l=0},\ref{eq:Ftl=0}) is the reality of expressions on the left hand side, 
and we take the more strict local reality at each point of invariant mass, 
{\small
\beq
\mathrm{Im} \Big[ \varGamma^{s \ast}_K(s ) F_{0/t}^{(l=0)}(q^2,s)\Big] = 0 \,.
\label{eq:phase-zero}
\eeq}
To fulfil this requirement, a strong phase $\phi_{f_0}$ is introduced to compensate the phase difference 
between the kaon form factor and the modeled $\bar{B}_s \to KK$ form factors. 
Generally speaking, $\phi_{f_0}$ should depend on both the two variables $s$ and $q^2$, 
while in the single $f_0$ model the $q^2-$dependence disappears,
{\small
\beq
\delta_{\varGamma_K^s}(s) - \phi_{f_0}(s) = \mathrm{Arg}  \, \Big[ \frac{g_{f_0 KK}} {m_{f_0}^2 -s - i \sqrt{s} \Gamma_{f_0}(s)} \Big] \,. 
\label{eq:phase-zero-1}
\eeq}

The simple model in Eqs.(\ref{eq:ff-b2dipi-model-1}-\ref{eq:ff-b2dipi-model-3}) is inspired by the physics that 
the sum rules obtained for the $\bar{B}_s \to KK$ form factors, in the narrow width approximation, 
should reproduce the sum rules for the form factors of $\bar{B}_s \to f_0$ transition. 
To check this, let's consider the energy-dependent width of $f_0$ with including the loop effects of two kaons coupling, 
%\footnote{We discuss in the general formula without 
%specifying detail expression of $\Gamma_{f_0}(s)$, i.e., the Flatt\'e model \cite{FlatteRZ}.} \cite{GokalpNY}:
{\small
\beq
\Gamma_{f_0}(s) &=& \frac{g_{f_0 KK}^2 \beta_K(s)}{4 \sqrt{2} \pi \sqrt{s} } \Theta (s - 4m_K^2) \, \non
&=& \Gamma^{\mathrm{tot}}_{f_0} \frac{\beta_K(s)}{\beta_K(m_{f_0})} \frac{m_{f_0}}{\sqrt{s}} \Theta (s - 4m_K^2) \,.
\label{eq:f0pipi-width}
\eeq}
The width of $f_0$ is usually parameterized under the Flatt$\acute{e}$ model \cite{FlatteRZ,FlatteXV} 
with considering the location of $f_0$ in the invariant mass, say, below or above the threshold. 
While in the $\bar{B}_s$ decays, the case is different because the invariant mass is alway above the threshold\footnote{
Another reason for us not using the Flatt$\acute{e}$ model is that it gives a smaller value of $I^{f_0}=4.44^{+0.81}_{-0.98}$, 
which is close to $I^{f'_0}$ and damages the contribution hierarchy from different resonances, as we would see in Eqs.(\ref{eq:LCSR-B2pipi-2},\ref{eq:Ifs}).}, 
then we can take the conventional form of the width as described in Eq.\ref{eq:f0pipi-width}, 
In the single $f_0$ model, the $s$ flavoured kaon form factor is written as 
{\small
\beq
\frac{\varGamma_K^{s \ast}(s)}{m_s} \Big\vert_{f_0} = \frac{g_{f_0 KK} m_{f_0} \bar{f}^s_{f_0} }{m_{f_0}^2 - s + i \sqrt{s} \Gamma_{f_0}(s)} 
\, e^{-i\phi_{f_0}(s,q^2)} \,.
\label{eq:pipi-ff-f0}
\eeq}
Substituting Eqs.(\ref{eq:ff-b2dipi-model-1},\ref{eq:pipi-ff-f0}) into Eq.\ref{eq:F0l=0}, %and taking into account Eq.\ref{eq:f0pipi-width}, 
the left hand side of Eq.\ref{eq:F0l=0} becomes 
{\small
\beq
&~& m_{f_0} \bar{f}_{f_0}\, \mathcal{F}^{\bar{B}_s \to f_0}_+(q^2) \int^{s^{2K}_0}_{4m_K^2} ds e^{-s/M^2} 
\frac{1}{\pi} \frac{\Gamma_{f_0}(s) \sqrt{s} }{(m_{f_0}^2-s)^2+ s \Gamma_{f_0}^2(s)}  \non
&~&\xrightarrow{\Gamma_{f_0}^{\mathrm{tot}} \rightarrow 0} 
m_{f_0} \bar{f}^s_{f_0} \, \mathcal{F}_+^{\bar{B}_s \to f_0}(q^2) \, e^{- m_{f_0}^2/M^2} \,.
\label{eq:f0-pole}
\eeq}
Similarly, we can reproduce the LCSRs for form factors $\mathcal{F}^{\bar{B}_s \to f_0}_{-}$ and $\mathcal{F}^{\bar{B}_s \to f_0}_0$ 
with taking into account the resonance models in Eq.\ref{eq:ff-b2dipi-model-2} and Eq.\ref{eq:ff-b2dipi-model-3}, respectively. 
What's more, the relation defined in Eq.\ref{eq:relation-f0-f+} also holds in the resonance models.

\subsection{Numerics}\label{sub:numerics}

We employ the $z-$series expansion for heavy-to-light transition form factors \cite{BourrelyZA}, with $j=+,-,0$,
{\small
\beq
\mathcal{F}^{\bar{B}_s \to f_0}_{j}(q^2) = 
\frac{\mathcal{F}^{\bar{B}_s \to f_0}_j(0)}{1-q^2/m_{B_s}^2} \Big\{1 + b_{\mathcal{F}_j} \,\zeta(q^2) + c_{\mathcal{F}_j}  \, \zeta^2(q^2) \Big\} \,.
\label{eq:z-para}
\eeq }
$\mathcal{F}^{\bar{B}_s \to f_0}_j(0)$ is the value at the full recoiled energy, 
the parameters $b_{\mathcal{F}_j}, c_{\mathcal{F}_i}$ indicate the coefficients associated with the $\zeta-$functions, 
{\small
\beq
&&\zeta(q^2) = z(q^2) - z(0)  \,,
\label{eq:z-para-zeta} \\
&&z(q^2) = \frac{\sqrt{t_+ - q^2} - \sqrt{t_+ - t_0}}{\sqrt{t_+ - q^2} + \sqrt{t_+ - t_0}} \,,
\label{eq:z-para-z}
\eeq}
with the definitions $t_\pm \equiv (m_{B_s} \pm m_{f_0})^2$ and $t_0 \equiv t_+ (1- \sqrt{1-t_-/t_0})$. 

The $S-$wave $\bar{B}_s \to KK$ form factors, under the single $f_0$ model, is rearranged in a general formula as  
{\small
\beq
&~&\Big[X_{\mathcal{F}_j} \, I(s_0^{2K}, M^2, \Gamma_{f_0}^{\text{tot}})  \Big]\, 
\frac{\kappa_{\mathcal{F}_j}+ \eta_{\mathcal{F}_j} \zeta(q^2) + \rho_{\mathcal{F}_j} \zeta^2(q^2)}{1-q^2/m_{B_s}^2} \, \non
&=& I^{\mathrm{OPE}}_j (s_0^{2K,} M^2, q^2)  \,,
\label{eq:LCSR-B2pipi}
\eeq}
where for the sake of brevity we introduce the following notations:
{\small
\beq
&&\kappa_{\mathcal{F}_j} \equiv \vert g_{f_0 KK} \vert \mathcal{F}^{\bar{B}_s \to f_0}_j(0) \,, \non
&&\eta_{\mathcal{F}_j} \equiv b_{\mathcal{F}_j}  \vert g_{f_0 KK} \vert \mathcal{F}^{\bar{B}_s \to f_0}_j(0) \,,  \non
&&\rho_{\mathcal{F}_j} \equiv  c_{\mathcal{F}_j}  \vert g_{f_0 KK} \vert \mathcal{F}^{\bar{B}_s \to f_0}_j(0) \,, \non
&&X_{\mathcal{F}_+} = X_{\mathcal{F}_-} = 1 \,,\,\,\,\,\,\,X_{\mathcal{F}_0} = (m_{B_s}^2-m_{f_0}^2) \,.
\label{eq:LCSR-B2pipi-notation}
\eeq}
The integral coefficient on the left hand side reads as
{\small
\beq
&~& I(s_0^{2K}, M^2, \Gamma_{f_0}^{\mathrm{tot}}) \non
&=& \frac{1}{4 \sqrt{2} \pi^2} \int_{4m_K^2}^{s_0^{2K}} ds e^{-s/M^2} 
\frac{\beta_K(s) \vert \varGamma^s_K(s)/m_s\vert }{\sqrt{(m_{f_0}^2-s)^2 + s \Gamma_{f_0}^2(s)}} \,.
\label{eq:LCSR-B2pipi-inter}
\eeq}
$I^{\mathrm{OPE}}_j$ represents the OPE calculations on the right hand side of Eqs.(\ref{eq:F0l=0},\ref{eq:Ft0l=0},\ref{eq:Ftl=0}). 
There is no physical requirement that the threshold value $s_0^{2K}$ should be equal to $s_0$, 
we fixed it in an independent way by considering the correlation function in Eq.\ref{eq:corr-2psr} with $KK$ interpolating. 
The 2pSRs is then written in terms of the scalar isoscalar kaon form factor, 
{\small
\beq
\int_{4m_K^2}^{s_0^{2K}} ds \, e^{-s/M^2} \, \frac{\beta_K(s)}{8\pi^2}  \Big\vert \frac{ \varGamma^s_K(s)}{m_s}\Big\vert^2 
= \Pi^{\mathrm{OPE}}_{\text{2pSRs}}(s_0^{2\pi},M^2) \,.
\label{eq:2psr-2pi}
\eeq}
We then determine the value $s_0^{2K} = 2.0$ GeV$^2$,  closing to it taken in the case of single meson interpolating.  

Besides the bound state $f_0$, it is nature to question what's the roles of the excited states $f'_0,f''_0$ in the $\bar{B}_s \to KK$ transition\footnote{
Hereafter we take $f_0$, $f'_0$ and $f''_0$ to denote the meson states $f_0(980)$, $f_0(1500)$ and $f_0(1710)$, respectively. 
We do not consider $f_0(1370)$ as a separated resonance due to the weak coupling of $f_0(1370)$ to $KK$ state.}.  
To include these effects, we suggest the $f_0+f'_0+f''_0$ model by appending $f'_0$ and $f''_0$ states to Eq.(\ref{eq:ff-b2dipi-model-3}),  
{\small
\beq
&~&\frac{1}{\sqrt{2}} F_t^{(l=0)}(s,q^2) \sqrt{q^2} \non 
&=& \sum_{S=f_0,f'_0,f''_0} 
\frac{g_{S KK} \, \mathcal{F}^{\bar{B}_s \to S}_0(q^2) \, (m_{B_s}^2- m_{S}^2) }{m_{S}^2 - s - i \sqrt{s} \Gamma_{S}(s)}  \, e^{i\phi_{S}(s)} \,.
\label{eq:ff-b2dipi-model-4} 
\eeq
}
We tacitly assume that the strong phase $\phi_S$ associated to each intermediate state is only dependent on the invariant mass of dikaon state, 
which means we do not consider the interaction effect between different resonances 
when introducing the strong phases to satisfy Eq.\ref{eq:phase-zero}.  
By this way, the $\bar{B}_s \to S$ form factors for different intermediate resonances are linear to each other: 
$\mathcal{F}_j^{f'_0}(q^2)= \mathcal{\gamma}^{f'_0}_{\mathcal{F}_j} \mathcal{F}_j^{f_0}(q^2)$ and 
$\mathcal{F}_j^{f''_0}(q^2)= \mathcal{\gamma}^{f''_0}_{\mathcal{F}_j} \mathcal{F}_j^{f_0}(q^2)$. 
The dimensionless parameters $\mathcal{\gamma}^{f'_0}_{\mathcal{F}_j}$ and $\mathcal{\gamma}^{f''_0}_{\mathcal{F}_j}$ 
indicate the relative size of $\bar{B}_s \to f'_0$ and $f''_0$ form factors comparing to the $\bar{B}_s \to f_0$ form factors, respectively. 
The LCSRs in Eq.\ref{eq:LCSR-B2pipi}, in the case of three resonance model, is modified to 
{\small
\beq
&~& \sum_{S=f_0,f'_0,f''_0} \Big[ \mathcal{\gamma}^S_{\mathcal{F}_j} \, X^S_{\mathcal{F}_j} \, I^S%(s_0^{2\pi}, M^2, \Gamma_S^{\text{tot}}) 
\Big]  \frac{\kappa_{\mathcal{F}_j}+ \eta_{\mathcal{F}_j} \zeta(q^2) + \rho_{\mathcal{F}_j} \zeta^2(q^2)}{1-q^2/m_{B_s}^2} \non
&=& I^{\mathrm{OPE}}_j (s_0^{2K,} M^2, q^2) \,.
\label{eq:LCSR-B2pipi-2}
\eeq}
We quote the values of the integral coefficients (in unit of $10^{-2}$)
\beq
I^{f_0} =  9.54^{+1.48}_{-1.54}  \,, \,\,\,
I^{f'_0} = 3.67^{+2.30}_{-1.38}    \,,\,\,\, 
I^{f''_0} = 1.82^{+0.73}_{-0.57} \,.\,\,\,\,\,\,
\label{eq:Ifs}
\eeq
In these integrals with taking the bound limit at $4 m_K^2$, only the right half of the peaking region in figure {\ref{fig:2}} is taken into account, 
so the relative sizes of integral coefficients $I^{f'_0}, I^{f''_0}$ to $I^{f_0}$ ($I^{f'_0}/I^{f_0},I^{f''_0}/I^{f_0}$) can be expected to be double of that in the $B \to \pi\pi$ case \cite{ChengSMJ}. 
 
%%-----------------------------------------------------------------------
\begin{table}[t]
\caption{{\footnotesize The fitting result for $\mathcal{F}_i$ in $f_0+f'_0+f''_0$ model.}}
\begin{center}
\begin{tabular}{c|c|c}
\toprule[1pt] 
$\vert g_{S \pi\pi} \vert \mathcal{F}_j$ & $\vert g_{S \pi\pi} \vert \mathcal{F}_+ $ &  $\vert g_{S \pi\pi} \vert \mathcal{F}_0 $ \\
\midrule[0.5pt]
$\kappa^{f_0}_{\mathcal{F}_j} \, (\mathrm{GeV})$ & $0.56^{+0.02}_{-0.08} $ & $0.40^{+0.01}_{-0.05} $ \non
$\eta^{f_0}_{\mathcal{F}_j} \, (\mathrm{GeV})$  &  $-0.97^{-0.01}_{+0.13}$ & $0.19^{+0.15}_{-0.19}$ \non
$\rho^{f_0}_{\mathcal{F}_j}\, (\mathrm{GeV})$ & $-12.5^{+2.60}_{-1.42}$ & $0.61^{-0.60}_{+0.57}$ \non
$\mathcal{\gamma}^{f'_0}_{\mathcal{F}_j}$ & $0.49^{+0.45}_{-0.17}$ & $0.48^{+0.26}_{-0.16}$ \non
$\mathcal{\gamma}^{f''_0}_{\mathcal{F}_j}$ & $0.66^{+0.80}_{-0.20}$ & $0.62^{+0.58}_{-0.26}$ \non
\midrule[0.4pt]
\midrule[0.4pt]
& $\mathcal{F}_+^{\bar{B}_s \to f_0}(0)$ & $\mathcal{F}_0^{\bar{B}_s \to f_0}(0)$ \non
\midrule[0.5pt]
& $0.52 \pm 0.10 $ & $0.37 \pm 0.06 $    \non
\bottomrule[1pt] 
\end{tabular}
\end{center}
\label{tab2}
\end{table}
%%-----------------------------------------------------------------------

The fitting result in the $f_0+f'_0+f''_0$ model are presented in table \ref{tab2},  
where the errors come from the sum rules parameters. 
For the total widths of the intermediate resonances we choose $\Gamma_{f_0}^{\mathrm{tot}} = 0.055$ GeV, 
$\Gamma^{\mathrm{tot}}_{f'_0} = 0.112$ GeV and $\Gamma^{\mathrm{tot}}_{f''_0} = 0.123$ GeV \cite{TanabashiOCA}. 
We note that varying the width of $f_0$ in $[0.01, 0.1]$ GeV brings another uncertainty to $I^{f_0}$ by $\vert^{+2.55}_{-1.40}$, 
while the width effects of $f'$ ($f''$) to $I^{f'_0}$ ($I^{f''_0}$) is negligible since their widths are much smaller. 
In the fit we also use the condition at $q^2 = 0$
{\small
\beq
\sum_{S=f_0,f'_0,f''_0} \Big[ \mathcal{\gamma}^S_{\mathcal{F}_j} \, X^S_{\mathcal{F}_j} \, I^S%(s_0^{2\pi}, M^2, \Gamma_S^{\text{tot}}) 
\Big]  \kappa_{\mathcal{F}_j} = I^{\mathrm{OPE}}_j (s_0^{2K,} M^2, 0) \,,
\eeq} 
and also the hierarchy anstz of different resonances in the left hand side of Eq.\ref{eq:LCSR-B2pipi-2}, 
say, $0 < \mathcal{\gamma}^{f'}_{\mathcal{F}_j}, \mathcal{\gamma}^{f''}_{\mathcal{F}_j} <1$. 
At the bottom of Tab.\ref{tab2}, for the comparison we supplement the $\bar{B}_s \to f_0$ form factors 
calculated in section \ref{LCSRs} under the narrow width approximation. 
In principle, the strong coupling $|g_{f_0 KK}|$ can be extracted out 
in case we have the reliable prediction for $\bar{B}_s \to f_0$ form factors, $\mathcal{F}_+$ and $\mathcal{F}_0$. 
With the result obtained in the narrow width approximation as we demonstrated in the last section, 
we can estimate $|g_{f_0 KK}| = 1.08^{+0.05}_{-0.14}$ GeV, 
but keep in mind that the width/non-resonant effect is sizeable and we reserve another $\sim 50 \%$ uncertainty. 

%%-----------------------------------------------------------------------
\begin{figure}[t]
\resizebox{0.45\textwidth}{!}{
\includegraphics{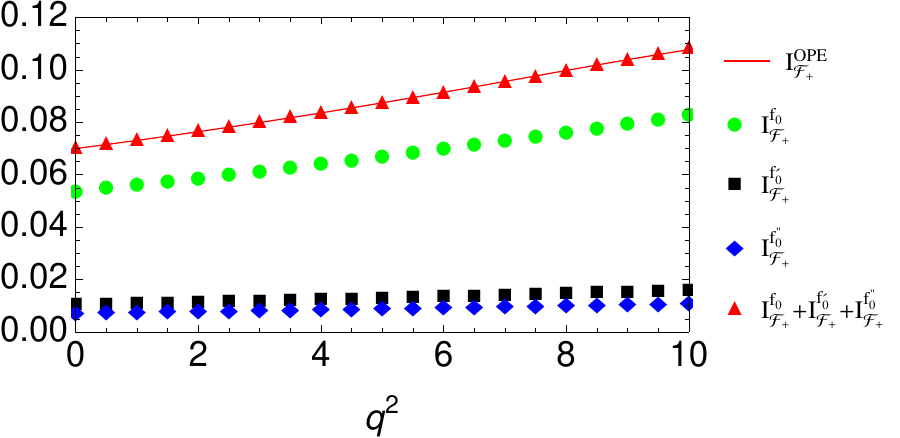}}\non
\resizebox{0.45\textwidth}{!}{
\includegraphics{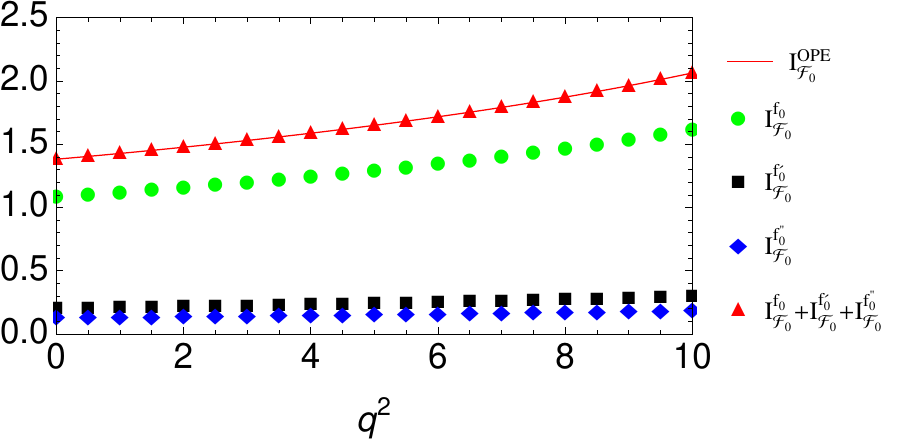}
}
\caption{The contributions to the OPE result $I^{\text{OPE}}_{\mathcal{F}_{+}}$ (up) 
and $I^{\text{OPE}}_{\mathcal{F}_{0}}$ (down) in the $f_0+f'_0+f''_0$ model.}
\label{fig:3}       
\end{figure}
%%-----------------------------------------------------------------------

\begin{figure}
\resizebox{0.45\textwidth}{!}{
\includegraphics{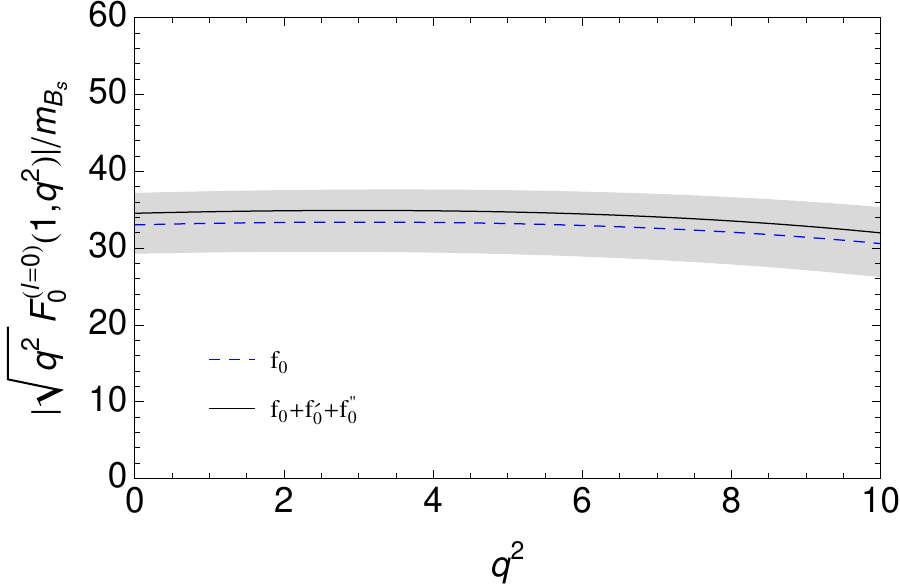}}\non
\resizebox{0.45\textwidth}{!}{
\includegraphics{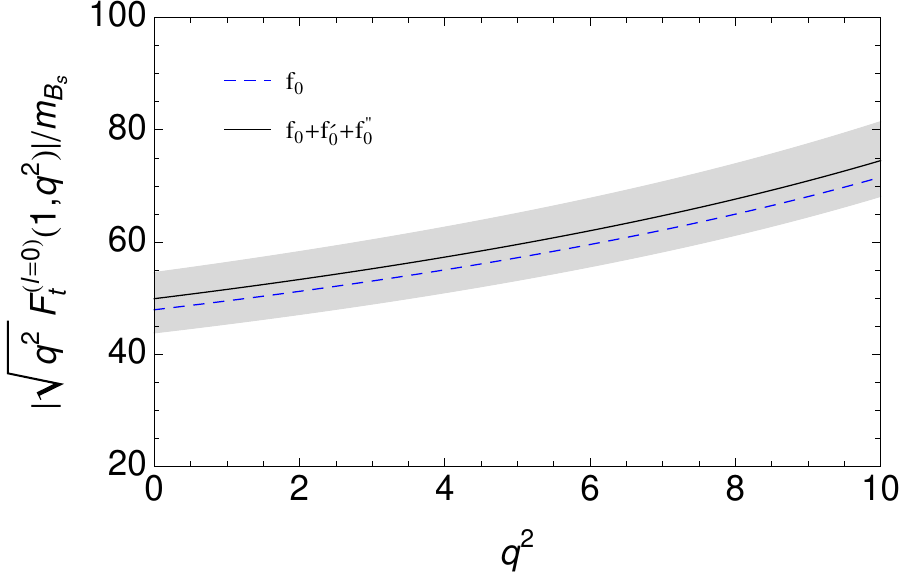}
}
\caption{$\sqrt{q^2} F_{0}^{(l=0)}(1,q^2)/m_{B_s}$ and $\sqrt{q^2} F_{t}^{(l=0)}(1,q^2)/m_{B_s}$ obtained under the $f_0+f'_0+f''_0$ model.}
\label{fig:4}       
\end{figure}

The contribution from each resonance to the OPE result (Eq.\ref{eq:LCSR-B2pipi-2}) is listplotted in Fig.\ref{fig:3}, 
from which the expected leading role of $f_0$ is confirmed. 
We plot in Fig.\ref{fig:4} for the $S-$wave $\bar{B}_s \to KK$ form factors in the $f_0+f'_0+f''_0$ model, 
for convenience we also show the part of contributions from $f_0$ in blue dashed curves. 
It is easy to see the overwhelming role of $f_0$, while the contributions from $f'_0$ and $f''_0$ account only $\sim 5\%$. 
The result at the full recoiled energy $\sqrt{q^2}F_t^{(l=0)}(1,0)/m_{B_s} = 54.0^{+4.0}_{-7.0}$ is much larger than 
the result for $S-$wave $B \to \pi\pi$ form factors obtained in the LCSRs with $2\pi$DAs \cite{ChengHPQ,HambrockAOR,ChengSFK}, 
with the asymptotic prediction $\sqrt{q^2}F_{t,\mathrm{asy}}^{(l=0)}(4m_\pi^2,0)/m_B = 5.40 \pm 1.00$, 
this discrepancy is explained by the strong threshold effect of $f_0$ in the $\bar{B}_s \to KK$ decay.

%%-----------------------------------------------------------------------
%%-----------------------------------------------------------------------

\section{Conclusion}\label{conclusion}

In this paper we calculate the $\bar{B}_s \to f_0(980)$ form factor from the light-cone sum rules with $B-$meson DAs, 
and investigate the $S-$wave $\bar{B}_s \to KK$ form factors to study the width effect, 
basing on the assumption that $f_0$ is dominated by the $\bar{s}s$ configuration. 
With taking the conventional quark-antiquark assignment, we revisit the 2pSRs for the mass and decay constant of $f_0(980)$. 
For the $\bar{B}_s \to f_0(980)$ form factors, 
we find that the high twist two-particle and the three-particle $B-$meson DAs 
give $25\%$ correction separately, and their total correction to certain form factors can be about $50\%$. 
In order to investigate the width effect, 
we suggest the three resonance states model to parameterize the $S-$wave $\bar{B}_s \to KK$ form factors, 
the fitting result shows the dominant role of $f_0$, 
and as a by-product, suggest a new way to determine the strong coupling $\vert g_{f_0KK} \vert$. 
The residual uncertainty of our prediction mainly comes from the freedom to choose the widths of $f_0$. 
%Further improvements on this project include mainly the follows:  
%(i) pushing the OPE calculation to the next-to-leading-order of QCD radiation to improve the theoretical accuracy, 
%(ii) studying the $S-$wave $\bar{B}_s \to KK$ form factors from the LCSRs with dikaon DAs, 
%of course we should first establish the wave function of dikaon system, 
%(iii) and last but not least, forwarding this approach to calculate the $B \to K\pi$ form factors which is more interested in phenomenology.

\section*{Acknowledgments}
We are grateful to Hai-yang Cheng and Ling-yun Dai for helpful discussions. 
This work is supported by the National Science Foundation of China under Grant No.118050 60 and No.11905056, 
"the Fundamental Research Funds for the Central Universities" under Grant No.531118010176 and No.531118010258. 
S. C. is grateful to the High energy theory group at Institute of Physics, Academia Sinica for hospitality and for financial
support where this work was finalized.

\begin{appendix}

\section{$B$ meson LCDAs}\label{app:B-LCDAs}

Several models have been suggested for the LCDAs with definite twists, 
which incorporate the correct low-momentum behaviour and satisfy the (tree-level) equation of motion (EOM) constraints \cite{GrozinPQ,BraunLIQ,LuCFC}. 
In \cite{BenekeWJP}, a more general ansatz is proposed with comprising all of these models as particular cases.

For the two-particle $B_s$ meson DAs demonstrated in Eq.\ref{eq:B-WFs-1},  
$\phi_+(\omega)$ and $\phi_-(\omega)$ are the leading and subleading twist DAs, 
and $g_+(\omega)$ and $g_-(\omega)$ are DAs at twist-4 and twist-5, respectively. 
The general model are quoted as \cite{BenekeWJP}
{\small
\beq
&&\phi_+(\omega) = \omega f(\omega) \,, 
\label{eq:2p-B-LCDAs-3}\\
&&\phi_-(\omega) =F(\omega)  \non
&&\hspace{1cm}- \frac{1}{6} \varkappa (\lambda_E^2-\lambda_H^2) \Big[ \omega^2 f'(\omega) + 4 \omega f(\omega) -2 F(\omega) \Big] \,, 
\label{eq:2p-B-LCDAs-3}\\
&&g_+(\omega) \simeq g_+^{WW}(\omega) = \frac{1}{8} \int_\omega^\infty \Big[ \omega^2 + 3\rho^2 - 4\bar{\mathrm{\Lambda}} \rho \Big] f(\rho) \,, 
\label{eq:2p-B-LCDAs-5}\\
&&g_-(\omega) = -\frac{3 \omega}{4} \mathcal{F}(\omega)  \non
&&\hspace{1cm} - \frac{\omega}{4} \varkappa (\lambda_E^2-\lambda_H^2) 
\Big[ \frac{\omega^3}{3} f(\omega) + \omega F(\omega) - \mathcal{F}(\omega) \Big] \,. 
\label{eq:2p-B-LCDAs-6}
\eeq}
The general function $f(\omega)$  is normalized as $\int_0^\infty d\omega \omega f(\omega) = 1$ 
and decreases sufficiently fast at $\omega \to \infty$. 
In Eqs.(\ref{eq:2p-B-LCDAs-3},\ref{eq:2p-B-LCDAs-6}), three auxiliary functions are introduced to simplify the expression, 
{\small
\beq
&&f'(\omega) = \frac{df(\omega)}{d\omega} \,,\\
&&F(\omega) \equiv \int_\omega^\infty d\rho f(\rho) \,, \\
&&\mathcal{F}(\omega) \equiv \int_\omega^\infty d\rho_2 \int_{\rho_2}^\infty d\rho_1 f(\rho_1) \,.
\eeq}
The normalization constant $\varkappa$ is determined from the leading twist function $f(\omega)$ via the EOM relations
{\small
\beq
&&\int_0^\infty d\omega \omega \phi_+(\omega) = \frac{4}{3}\bar{\mathrm{\Lambda}} \,, 
\label{eq:EOM-1}\\
&&\int_0^\infty d\omega \omega^2 \phi_+(\omega) = 2 \bar{\mathrm{\Lambda}}^2 + \frac{1}{3}(2\lambda_E^2+\lambda_H^2) \,,
\label{eq:EOM-2}\\
&&\varkappa^{-1} = \frac{1}{6} \int_0^\infty d\omega \omega^3 \phi_+(\omega) = 
\bar{\mathrm{\Lambda}}^2 + \frac{1}{6}(2\lambda_E^2+\lambda_H^2) \,. 
\label{eq:EOM-3}
\eeq}

Concerning the three-particle $B_s$ meson DAs, the definitions in Eq.\ref{eq:B-WFs-2} by Lorentz structures should not be confused with 
the definitions by means of definite twists, and they have the following relations \cite{BraunLIQ}
{\small
\beq
&&\Psi_A(\omega,\zeta) = \frac{1}{2} \Big[ \phi_3(\omega, \zeta) + \phi_4(\omega, \zeta) \Big] \,, \non
&&\Psi_V(\omega, \zeta) = \frac{1}{2} \Big[ -\phi_3(\omega, \zeta) + \phi_4(\omega, \zeta) \Big] \,,\non
&&X_A(\omega, \zeta) = \frac{1}{2} \Big[ -\phi_3(\omega, \zeta) - \phi_4(\omega, \zeta) +2 \psi_4(\omega, \zeta) \Big] \,, \non
&&Y_A(\omega, \zeta) = \frac{1}{2} \Big[ -\phi_3(\omega, \zeta) - \phi_4(\omega, \zeta) + \psi_4(\omega, \zeta) - \psi_5(\omega, \zeta) \Big] \,,\non
&&\widetilde{X}_A(\omega, \zeta) = \frac{1}{2} \Big[ -\phi_3(\omega, \zeta) + \phi_4(\omega, \zeta) - 2\widetilde{\psi}_4(\omega, \zeta) \Big] \,, \,\non
&&\widetilde{Y}_A(\omega, \zeta) = \frac{1}{2} \Big[ -\phi_3(\omega, \zeta) + \phi_4(\omega, \zeta) 
- \widetilde{\psi}_4(\omega, \zeta) + \widetilde{\psi}_5(\omega, \zeta) \Big] \,, \non
&&W(\omega, \zeta) = \frac{1}{2} \Big[ \phi_4(\omega, \zeta) - \psi_4(\omega, \zeta) - \widetilde{\psi}_4(\omega, \zeta)  \non
&& \hspace{1.7cm} + \widetilde{\phi}_5(\omega, \zeta) + \psi_5(\omega, \zeta) + \widetilde{\psi}_5(\omega, \zeta) \Big] \,, \non
&&Z(\omega, \zeta) = \frac{1}{4} \Big[ -\phi_3(\omega, \zeta) + \phi_4(\omega, \zeta) - 2 \widetilde{\psi}_4(\omega, \zeta) \non
&& \hspace{1.7cm} + \widetilde{\phi}_5(\omega, \zeta) + 2 \widetilde{\psi}_5(\omega, \zeta) - \phi_6(\omega, \zeta) \Big] \,.
\label{eq:relation-LCDAs}
\eeq}
The general model for the DAs with definite twists are \cite{BenekeWJP}
{\small
\beq
&&\phi_3(\omega, \zeta) = -\frac{1}{2} \, \varkappa \, (\lambda_E^2-\lambda_H^2) \, \omega \, \zeta^2 \, f'(\omega+\zeta) \,,\\
&&\phi_4(\omega, \zeta) = \frac{1}{2} \, \varkappa \, (\lambda_E^2+\lambda_H^2) \, \zeta^2 \, f(\omega+\zeta) \,, \\
&&\psi_4(\omega, \zeta) = \varkappa \, \lambda_E^2 \, \omega \, \zeta \, f(\omega+\zeta) \,, \\
&&\widetilde{\psi}_4(\omega, \zeta) =  \varkappa \, \lambda_H^2 \, \omega \, \zeta \, f(\omega+\zeta) \,, \\
&&\widetilde{\phi}_5(\omega, \zeta) = - \varkappa \, (\lambda_E^2+\lambda_H^2) \, \omega \, F(\omega+\zeta) \,, \\
&&\psi_5(\omega, \zeta) = \varkappa \, \lambda_E^2 \, \zeta \, F(\omega+\zeta) \,, \\
&&\widetilde{\psi}_5(\omega, \zeta) = \varkappa \, \lambda_H^2 \, \zeta \, F(\omega+\zeta) \,, \\
&&\phi_6(\omega, \zeta) = - \varkappa \, (\lambda_E^2-\lambda_H^2) \mathcal{F}(\omega+\zeta) \,.
\label{eq:3p-B-LCDAs}
\eeq}
$\lambda_E$ and $\lambda_H$ are the parameters entered in the normalization conditions 
{\small
\beq
\Psi_A(x=0) = \frac{\lambda_E^2}{3} \,,\,\,\,\,\,\,
\Psi_V(x=0) = \frac{\lambda_H^2}{3} \,.
\label{eq:norm-phiAV}
\eeq}

It is known that the EOM, as shown in Eqs.(\ref{eq:EOM-1}-\ref{eq:EOM-3}), 
imply the connections between the two-particle and three-particle LCDAs, 
{\small 
\beq
&&\omega_0 = \lambda_B = \frac{2}{3}\bar{\mathrm{\Lambda}} \,, \,\,\, 
2\bar{\mathrm{\Lambda}}^2 = 2\lambda_E^2 + \lambda_H^2 \,, \,\,\,\,\,\, \mathrm{Exp-Model} \,, 
\label{eq:Exp-EOM}\\
&&\omega_0 = \frac{5}{2} \lambda_B = 2 \bar{\mathrm{\Lambda}} \,, \,\,\,
\bar{\mathrm{\Lambda}}^2 = 2\lambda_E^2 + \lambda_H^2 \,, \,\,\,\,\,\, \mathrm{LD-model} \,,
\label{eq:LD-EOM}
\eeq}
with taking the general functions
{\small
\beq
&&f(\omega) = \frac{1}{\omega_0^2} e^{-\omega/\omega_0} \,,\,\,\,\,\,\hspace{2.2cm} \mathrm{Exp-Model}\,, \\
&&f(\omega) = \frac{5}{8\omega_0^5} (2\omega_0 - \omega)^3 \Theta[2\omega_0-\omega] \,,\,\,\,\,\, \mathrm{LD-Model}\,.  
\eeq}
The normalization constants in these two particular models are 
{\small 
\beq
&&\varkappa = \frac{1}{3\omega_0^2} \,, \,\,\,\,\,\, \mathrm{Exp-Model} \,, 
\label{eq:kappa-Exp}\\
&&\varkappa = \frac{7}{2\omega_0^2} \,, \,\,\,\,\,\,\mathrm{LD-Model}\,.  
\eeq}
We use the exponential models in our numerical evaluation. 

\section{Coefficients in the three-particle correction}\label{app:coeff-3p}

%We collect here for the coefficients of three-particle $B-$meson DAs contributed to $B \to S$ form factors. 
%As what we did for the contributions from two-particle $B-$meson LCDAs, here we omit the light quark terms too.

\subsection{Correction coefficients to $\mathcal{F}_+(q^2)$} 
{\small
\beq
&&C_{1,\mathcal{F}_+}^{\,\Phi_A-\Phi_V}= - \frac{2u-2}{\bar{\sigma}m_B^2} \,, \non
&&C_{2,\mathcal{F}_+}^{\, \Phi_A-\Phi_V} = - \frac{(2u-2)(m_B^2-q^2) + (2u+1)\bar{\sigma}^2m_B^2}{\bar{\sigma}m_B^2} \,,\non
&& C_{2,\mathcal{F}_+}^{\, \Phi_V} =  - 6 u \bar{\sigma} \,,\non 
&&C_{2,\mathcal{F}_+}^{\, \overline{X}_A} = \frac{(2u-1)}{m_B} \,,\,\,\,\,\,\,
C_{3,\mathcal{F}_+}^{\, \overline{X}_A} = \frac{2(2u-1)(\bar{\sigma}^2m_B^2-q^2)}{m_B} \,, \non
&&C_{2,\mathcal{F}_+}^{\, \overline{Y}_A+\overline{W}} =  \frac{18}{m_B} \,, \,\,\,\,\,\,
C_{2,\mathcal{F}_+}^{\, \overline{\widetilde{X}}_A} = \frac{1}{m_B} \,, \,\,\,\,\,\,
C_{2,\mathcal{F}_+}^{\, \overline{\widetilde{Y}}_A} = - \frac{2}{m_B} \,, \non
&&C_{3,\mathcal{F}_+}^{\, \overline{\widetilde{X}}_A} 
= - 8 \bar{\sigma} \frac{\left[ 2 m_S^2 \bar{\sigma} + (1-2\sigma)(m_B^2-m_S^2-q^2)-2\sigma q^2\right]}{m_B}  \,, \non
&&C_{3,\mathcal{F}_+}^{\, \overline{\widetilde{Y}}_A} = - C_{3,\mathcal{F}_+}^{\, \overline{\widetilde{X}}_A}  \,,\non
&&C_{3,\mathcal{F}_+}^{\, \overset{=}{W}} = -16 \bar{\sigma} (u+u^2) - 4 (4u^2+u) \non
&& \hspace{1.2cm} \cdot \frac{\left[ 2 m_S^2 \bar{\sigma} + (1-2\sigma)(m_B^2-m_S^2-q^2)-2\sigma q^2\right]}{m_B^2} \,.  
\label{eq:ff-3p-C-f+} 
\eeq}
\subsection{Correction coefficients to $\mathcal{F}_-(q^2)$}
{\small
\beq
&&C_{1,\mathcal{F}_-}^{\, \Phi_A-\Phi_V}= - \frac{2u-2}{\bar{\sigma}m_B^2} \,, \non
&&C_{2,\mathcal{F}_-}^{ \, \Phi_A-\Phi_V} = - \frac{(2u-2)(m_B^2-q^2) - (2u+1)\sigma\bar{\sigma}m_B^2}{\bar{\sigma}m_B^2} \,,\non
&&C_{2,\mathcal{F}_-}^{\, \Phi_V} = 6 u \sigma  \,,\non
&&C_{2,\mathcal{F}_-}^{\, \overline{X}_A} = - \frac{3(2u-1)}{m_B} \,,\non
&&C_{3,\mathcal{F}_-}^{\, \overline{X}_A} =  -\frac{2(2u-1)\sigma (\bar{\sigma}^2m_B^2-q^2)}{\bar{\sigma}m_B} \,, \non
&&C_{2,\mathcal{F}_-}^{\, \overline{Y}_A+\overline{W}} = \frac{18}{m_B} \,, \,\,\,\,\,\,
C_{2,\mathcal{F}_-}^{\, \overline{\widetilde{X}}_A} = \frac{1}{m_B} \,,\,\,\,\,\,\,
C_{2,\mathcal{F}_-}^{\, \overline{\widetilde{Y}}_A} = - \frac{2}{m_B} \,, \non
&&C_{3,\mathcal{F}_-}^{\, \overline{\widetilde{X}}_A} = 8 \sigma 
\frac{\left[ 2 m_S^2 \bar{\sigma} + (1-2\sigma)(m_B^2-m_S^2-q^2)-2\sigma q^2\right]}{m_B} \,, \non
&&C_{3,\mathcal{F}_-}^{\, \overline{\widetilde{Y}}_A} = - C_{3,\mathcal{F}_-}^{\, \overline{\widetilde{X}}_A}\,,\non
&&C_{3,\mathcal{F}_-}^{\, \overset{=}{W}} = 16 \sigma (u+u^2) -  4 (4u^2+u) \non
&& \hspace{1.2cm}\frac{\left[ 2 m_S^2 \bar{\sigma} + (1-2\sigma)(m_B^2-m_S^2-q^2)-2\sigma q^2\right]}{m_B^2} \,.  
\label{eq:ff-3p-C-f-} 
\eeq}
\subsection{Correction coefficients to $\mathcal{F}_{T,p}(q^2)$}
{\small
\beq
&&C_{2,\mathcal{F}_{T,p}}^{\, \Phi_A-\Phi_V} = - \frac{(2u-1)}{m_B} \,, \,\,\,\,\,\, 
C_{2,\mathcal{F}_{T,p}}^{\, \Phi_V} = -\frac{3u}{m_B} \,, \non 
&&C_{2,\mathcal{F}_{T,p}}^{\, \overline{X}_A} = 2 \frac{2u-1}{\bar{\sigma} m_B^2}  \,,\non
&&C_{3,\mathcal{F}_{T,p}}^{\, \overline{X}_A} = 
- 2 \frac{ [(2u-1) (\bar{\sigma}^2m_B^2-q^2) ]}{\bar{\sigma} m_B^2} \, \non
&&C_{2,\mathcal{F}_{T,p}}^{\, \overline{\widetilde{X}}_A} = - \frac{8}{m_B \sqrt{q^2}} \,, \non
&&C_{3,\mathcal{F}_{T,p}}^{\, \overline{\widetilde{X}}_A} = 
4 \bar{\sigma} \frac{[\bar{\sigma} (m_B^2-m_S^2-q^2) - 2 \sigma q^2 ]}{q^2} \,, \non
&&C_{3,\mathcal{F}_{T,p}}^{\, \overset{=}{W}} = 3 u \left[ \frac{\bar{\sigma}}{\sqrt{q^2}} - 
\frac{(m_B - \sqrt{q^2})}{m_B\sqrt{q^2} } \right] \,,\non
&&C_{3,\mathcal{F}_{T,p}}^{\, \overset{=}{Z}} = (24 u^2 + u -47) \non
&&\hspace{1.4cm} \cdot \frac{\left[\bar{\sigma} (m_B^2-m_S^2+q^2) - 2 (m_B-\sqrt{q^2}) \sqrt{q^2} \right]}{2 m_B q^2} \,. 
\label{eq:ff-3p-C-fT1} 
\eeq}

\subsection{Correction coefficients to $\mathcal{F}_{T,q}(q^2)$}
{\small
\beq
&&C_{1,\mathcal{F}_{T,q}}^{\, \Phi_A-\Phi_V} = \frac{2u-1}{\bar{\sigma}m_B} \, \non
&&C_{2,\mathcal{F}_{T,q}}^{\, \Phi_A-\Phi_V} = - \frac{(2u-1)[\bar{\sigma}^2m_B^2-q^2(1-2\sigma)]}{\bar{\sigma}m_B} \,, \non
&&C_{1,\mathcal{F}_{T,q}}^{\, \Phi_V} = \frac{3u}{\bar{\sigma}m_B} \,, \,\,\,\,\,\, 
C_{2,\mathcal{F}_{T,q}}^{\, \Phi_V} = -\frac{3u [\bar{\sigma}^2m_B^2-q^2(1-2\sigma)]}{\bar{\sigma}m_B} \,, \non
&&C_{1,\mathcal{F}_{T,q}}^{\, \overline{X}_A} = - \frac{2(2u-1)}{\bar{\sigma}^2 m_B^2} \,, \,\,\,\,\,\,
C_{2,\mathcal{F}_{T,q}}^{\, \overline{X}_A} = \frac{ 4 \sigma (2u-1) q^2}{\bar{\sigma}^2 m_B^2}  \,,\non
&&C_{3,\mathcal{F}_{T,q}}^{\, \overline{X}_A} = 
2 \frac{ [(2u-1) (\bar{\sigma}^2m_B^2-q^2) ][\bar{\sigma}^2m_B^2-q^2(1-2\sigma)]}{\bar{\sigma}^2 m_B^2} \,, \non
&&C_{2,\mathcal{F}_{T,q}}^{\, \overline{\widetilde{X}}_A} = 4 - \frac{16 \sqrt{q^2}}{m_B} \,,  \non 
&&C_{3,\mathcal{F}_{T,q}}^{\, \overline{\widetilde{X}}_A} = 8 \sigma [\bar{\sigma} (m_B^2-m_S^2-q^2) - 2 \sigma q^2 ] \,, \non
&&C_{2,\mathcal{F}_{T,q}}^{\, \overline{\widetilde{Y}}_A} = 120 \,,  \non
&&C_{3,\mathcal{F}_{T,p}}^{\, \overset{=}{W}} = 6 u \left[ \sigma \sqrt{q^2} + \frac{(m_B - \sqrt{q^2})\sqrt{q^2}}{m_B} \right] \,,\non
&&C_{3,\mathcal{F}_{T,p}}^{\, \overset{=}{Z}} = (24 u^2 + u -47) \non
&&\hspace{1.4cm} \cdot \frac{\left[\bar{\sigma} (m_B^2-m_S^2+q^2) - 2 (m_B-\sqrt{q^2}) \sqrt{q^2} \right]}{m_B} \,. 
\label{eq:ff-3p-C-fT2} 
\eeq}
\subsection{Correction coefficients to $\mathcal{F}_0(q^2)$}
{\small
\beq
&&C_{1,\mathcal{F}_0}^{\, \Phi_A-\Phi_V} = \frac{3(1 - 2u)}{2\bar{\sigma}m_B} \,, \non
&&C_{2,\mathcal{F}_0}^{\, \Phi_A-\Phi_V} = \frac{3(1 - 2u)}{2\bar{\sigma}m_B}(\bar{\sigma}^2m_B^2 -q^2)  \,, \non
&& C_{1,\mathcal{F}_0}^{\, \Phi_V} = - \frac{3u}{\bar{\sigma}m_B} \,, \,\,\,\,\,\,
C_{2,\mathcal{F}_0}^{\, \Phi_V} = - \frac{3u}{\bar{\sigma}m_B}(\bar{\sigma}^2m_B^2 -q^2) \,, \non
&&C_{1,\mathcal{F}_0}^{\,\overline{X}_A} = \frac{2u+1}{\bar{\sigma}^2m_B^2} \,, \non
&&C_{2,\mathcal{F}_0}^{\,\overline{X}_A} = (2u+7) - \frac{2(2u+1)(\bar{\sigma}^2m_B^2-q^2)}{\bar{\sigma}^2m_B^2}\,, \non
&&C_{3,\mathcal{F}_0}^{\,\overline{X}_A} = \frac{2u+1}{\bar{\sigma}^2m_B^2}(\bar{\sigma}^2m_B^2-q^2)^2 \,, \non
&&C_{2,\mathcal{F}_{0}}^{\, \overline{\widetilde{X}}_A} = \frac{3}{2} \,, \,\,\,\,\,
C_{3,\mathcal{F}_{0}}^{\, \overset{=}{W}} = 9 u \, \left[\bar{\sigma} (m_B - \sqrt{q^2}) - \sigma \sqrt{q^2} \right] \,, \non
&&C_{3,\mathcal{F}_{0}}^{\, \overset{=}{Z}} = - 24 u^2 \, \left[\bar{\sigma} (m_B - \sqrt{q^2}) - \sigma \sqrt{q^2} \right] \,. 
\label{eq:ff-3p-C-f0}
\eeq}

\end{appendix}

% BibTeX users please use
% \bibliographystyle{}
% \bibliography{}

\end{document}